\documentclass[10pt,twocolumn,twoside]{IEEEtran}
\usepackage{cite}

\usepackage{amsmath,graphicx}
\usepackage[utf8]{inputenc}
\usepackage{amssymb}  
\usepackage{amsthm}
\usepackage{amsfonts}
\usepackage{dsfont}
\usepackage{algorithm, algorithmic, verbatim}
\usepackage{etoolbox}
\newtoggle{showrevisions}

\newcommand{\R}{\mathds{R}}

\newcommand{\V}{\mathcal{V}}
\newcommand{\E}{\mathcal{E}}
\newcommand{\G}{\mathcal{G}}

\usepackage{graphicx}
\usepackage{color}
\usepackage{placeins}
\usepackage{float}
\usepackage{tabularx,colortbl}

\begin{document}
%
\title{Graph Transform Optimization with Application to Image Compression}

\author{Giulia Fracastoro, 
Dorina~Thanou,
and Pascal~Frossard
\thanks{G. Fracastoro is with the Department of Electronics and Telecommunications, 
Politecnico di Torino, Italy (e-mail: giulia.fracastoro@polito.it).}
\thanks{D. Thanou is with the Swiss Data Science Center, EPFL/ETHZ, Switzerland (e-mail: dorina.thanou@epfl.ch). }
\thanks{P. Frossard  is with the Signal Processing Laboratory (LTS4), EPFL, Lausanne, Switzerland (e-mail: pascal.frossard@epfl.ch).}
}
\maketitle

\begin{abstract}
In this paper, we propose a new graph-based transform and illustrate its potential application to signal compression. 
Our approach relies on the careful design of a graph that optimizes the overall rate-distortion performance through an effective graph-based transform. 
We introduce a novel graph estimation algorithm, which uncovers the connectivities between the graph signal values by taking into consideration the coding of both the signal and the graph topology  in rate-distortion terms. In particular, we introduce a novel coding solution for the graph  by treating the edge weights  as another graph signal that lies on the dual graph. Then, the cost of the graph description is introduced in the optimization problem by minimizing the sparsity of the coefficients of its graph Fourier transform (GFT) on the dual graph. In this way, we obtain a convex optimization problem whose solution defines an efficient transform coding strategy.  The proposed technique is a general framework that can be applied to different types of signals, and we show two possible application fields, namely natural image coding and piecewise smooth image coding. The experimental results show that  the proposed graph-based transform  outperforms  classical fixed transforms such as DCT for both natural and piecewise smooth images. In the case of depth map coding, the obtained results are even comparable to the state-of-the-art graph-based coding method, that are specifically designed for depth map images.
\end{abstract}

\IEEEpeerreviewmaketitle

\section{Introduction}
In the last years, the new field of signal processing on graphs has gained increasing attention \cite{shuman2013emerging}. Differently from classical signal processing, this new emerging field considers signals that lie on irregular domains, where the signal values are defined on the nodes of a weighted graph and the edge weights reflect the pairwise relationship between these nodes.  
Particular attention has been given to the design of  flexible graph signal representations, opening the door to new structure-aware transform coding techniques, and eventually to more efficient signal and image compression frameworks.  
As an illustrative example, an image can be represented by a graph, where the nodes are the image pixels and the edge weights capture the similarity between adjacent pixels.  Such a flexible representation permits to go beyond traditional transform coding by moving from classical fixed transforms such as the discrete cosine transform (DCT) \cite{ahm74} to graph-based transforms that are better adapted to the actual signal structure, such as the graph Fourier transform (GFT) \cite{hammond2011wavelets}. 
Hence, it is possible to obtain a more compact representation of an image, as the energy of the image signal is concentrated in the lowest frequencies. 
This provides a strong advantage compared to the classical DCT transform especially when the image contains arbitrarily shaped discontinuities. In this case, the DCT transform coefficients are not necessarily sparse and contain many high frequency coefficients with high energy. The GFT, on the other hand, may lead to sparse representations  and eventually more efficient compression.


However, one of the biggest challenges in graph-based signal compression remains the design of the graph and the corresponding transform. A good graph for effective transform coding should lead to easily compressible signal coefficients, at the cost of a small overhead for coding the graph. Most graph-based coding techniques focus mainly on images, and they construct the graph by considering pairwise similarities among pixel intensities \cite{shen2010edge,kim2012graph} or using a lookup table that stores the most popular GFTs \cite{hu2015multiresolution}. It has been shown that these methods could provide a significant gain in the coding of piecewise smooth images. Instead, in the case of natural images, the cost required to describe the graph often outweighs the coding gain provided by the adaptive graph transform, and often leads to unsatisfactory results. The  problem of designing a graph transform stays critical and may actually represent the major obstacle towards effective compression of signals that live on an irregular domain. 

In this work, we propose a new method for designing a graph transform that takes into account the coding of the signal values as well as the cost of coding the graph.  Second, we present an innovative way for  coding the graph by treating its edge weights as a graph signal that lies on the dual graph.  We then  compute the graph Fourier transform of this signal and code its quantized transform coefficients. 
The choice of the graph is thus posed as a rate-distortion optimization  problem. The cost of coding the signal is captured by minimizing the smoothness of the graph signal on the adapted graph. The coding cost of the graph itself is controlled by penalizing the sparsity of the graph Fourier coefficients of the edge weight signal that lies on the dual graph.   The solution of our optimization problem is a  graph that provides an effective tradeoff between the sparsity of the signal transform coefficients and the graph coding cost. 

We apply our method to two different types of signals, namely natural images and piecewise smooth images.
Experimental results on natural images confirm that the proposed algorithm can efficiently infer meaningful graph topologies, which eventually lead to a graph transform that can outperform non-adaptive classical transforms such as the DCT. 
Moreover,  we show that our method can significantly improve the classical DCT on piecewise smooth images, and it even leads to comparable results to the state-of-the-art graph-based depth image coding solutions. However, in contrary to these dedicated algorithms, it is important to underline that our framework is quite generic and can be applied to very different types of signals. 

Our work extends our previous study \cite{FracastoroTF16}, where we have shown that an initial version of our graph compression scheme could be successfully applied to compression of natural images. In this paper we show that the proposed method is highly general and it can be adapted to different types of signals. In addition, we present a deeper theoretical discussion on the GFT, highlighting its connection with the Karhunen-Loève transform and justifying its use for image coding.  The extension proposed in this paper offers a complete description of our novel system and a comprehensive performance analysis with extensive comparisons for various transforms.

The outline of the paper is as follows. We first discuss related work in Section II. We then introduce some preliminary definitions on graphs in Section III.  Next, we present the proposed graph construction problem in Section IV. The application of the proposed graph construction algorithm to image coding and the entire compression framework are described in Section V. Then, the experimental results on natural images and piecewise smooth images are presented in Section VI and VII, respectively. Finally we draw some conclusions in Section VIII.

\section{Related work}
\label{sec:rel_work}
In this section, we first provide a brief overview of transform coding. Then, we focus on graph-based coding and learning methods, that are closely related to the framework proposed in this work. 
\subsection{Transform coding}
Lossy image compression usually employs a 2D transform to produce a new image representation that lies in the transform domain \cite{sayood2012introduction}. Usually, the obtained transform coefficients are approximately uncorrelated and most of the information is contained in only a few of them. It is proved that the Karhunen-Loève transform (KLT) can optimally decorrelate a signal that has Gaussian entries \cite{goyal2000transform}. However, since the KLT is based on the eigendecomposition of the covariance matrix, this matrix or the transform itself has to be sent to the receiver. For this reason, the KLT is not practical in most circumstances \cite{sayood2012introduction}. The most common transform in image compression is the DCT \cite{ahm74}, which employs a fixed set of basis vector. It is known that the DCT is asymptotically equivalent to the KLT for signals that can be modelled as a first-order autoregressive process \cite{jain1979sinusoidal}. Nevertheless, this model fails to capture the complex and nonstationary behavior that is typically present in natural images. In the light of the above, transform design is still an active research field and in the last years many signal adaptive transforms have been presented. In this paper, we focus on a specific type of adaptive transforms, namely graph-based transforms.

\subsection{Graph-based image coding}
In the last years, graph signal processing has been applied to different image coding applications, especially for piecewise smooth images.
In \cite{kim2012graph,shen2010edge}, the authors propose a graph-based coding method where the graph is defined by considering pairwise similarities among pixel intensities. Another efficient graph construction method for piecewise smooth images has been proposed in \cite{hu2015multiresolution}, where the authors use a lookup table that stores the most popular graphs. Then, for each signal, they perform an exhaustive search choosing the best GFT in rate-distortion terms. Furthermore,  a new graph transform, called signed graph Fourier transform, has been presented in \cite{su2017graph}. This transform is targeted for compression of depth images and its underlying graph contains negative edges that describe negative correlations between pixel pairs. 

Recently, a number of methods using a graph-based approach have also been proposed for transform coding of inter and intra predicted residual blocks in video compression. A novel graph-based method for intra-frame coding has been presented in \cite{hu2015intra}, which introduces a new generalized graph Fourier transform. A graph-based method for inter predicted video coding has been introduced in \cite{egilmez2015graph}, where the authors design a set of simplified graph templates capturing the basic statistical characteristics of inter predicted residual blocks. Furthermore, a few separable graph-based transforms for residual coding have also been introduced. In \cite{egilmez2016gbst}, for example, the authors propose a new class of graph-based separable transforms for intra and inter predictive video coding. The proposed transform is based on two separate line graphs, where the edge weights are optimized using a graph learning problem. Another graph-based separable transform for inter predictive video coding has been presented in \cite{lu2016symmetric}. In this case, the proposed transform, called symmetric line graph transform, has symmetric eigenvectors and therefore it can be efficiently implemented.

Finally, a few graph-based methods have also been presented for natural image compression. In \cite{fracastoro2015predictive}, a new technique of graph construction targeted for image compression is proposed. This method employs innovative edge metrics, quantization and edge prediction technique. Moreover, in \cite{pavez2015gtt}, a new class of transforms called graph template transforms has been introduced for natural image compression, focusing in particular on texture images. Finally, a method for designing sparse graph structures that capture principal gradients in image code blocks is proposed in \cite{rotondo2015designing}. However, in all these methods, it is still not clear how to define a graph whose corresponding transform provides an effective tradeoff between the sparsity of the transform coefficients and the graph coding cost.

\subsection{Graph construction}
Several attempts to learn the structure and in particular a graph from data observations have been recently proposed, but not necessarily from a compression point of view. In \cite{pavez2016generalized,egilmez2016graph,pavez2017learning}, the authors formulate the graph learning problem as a precision matrix estimation with generalized Laplacian constraints. The same method is also used in \cite{egilmez2016gbst,lu2016symmetric}, where the authors use a graph learning problem in order to find the generalized graph Laplacian that best approximates residual video data. Moreover, in \cite{dong2014learning,kalofolias2016learn}, a sparse combinatorial Laplacian matrix is estimated from the data samples under a smoothness prior. 
Furthermore, in \cite{pavez2015gtt}, the authors use a graph template to impose on the graph Laplacian a sparsity pattern and approximate  the empirical inverse covariance based on that template. 

Even if all the methods presented above contain some constraints on the sparsity of the graph, none of them explicitly takes into account the real cost of representing and coding the graph. In addition, most of them do not really target images. Instead, in this paper, we go beyond prior art and we fill this gap by defining a new graph construction problem  that takes into account  the graph coding cost. Moreover, we show how our generic framework can be used for image compression.

\section{Basic definitions on graphs}
For any graph $\G=(\V,\E)$ where $\V$ and $\E$ represent respectively the node and edge sets with $|\V|=N$ and $|\E|=M$, we  define the weighted adjacency matrix $W\in \R^{N\times N}$ where $W_{ij}$ is the weight associated to the edge $(i,j)$ connecting nodes $i$ and $j$. For undirected graphs with no self loops, $W$ is symmetric and has null diagonal. The graph Laplacian is defined as $L=D-W$, where $D$ is a diagonal matrix whose $i$-th diagonal element $D_{ii}$ is the sum of the weights of all the edges incident to node $i$. Since $L$ is a real symmetric matrix, it is diagonalizable by an orthogonal matrix
\[
L=\Psi \Lambda \Psi^T,
\]
where $\Psi\in\mathbb{R}^{N\times N}$ is the eigenvector matrix of $L$ that contains the eigenvectors as columns, and $\Lambda\in\mathbb{R}^{N\times N}$ is the diagonal eigenvalue matrix, with eigenvalues sorted in ascending order.

In the next sections, we will use also an alternative definition of the graph Laplacian $L$ that uses the incidence matrix $B\in\mathbb{R}^{N\times M}$ \cite{gallier2014elementary}, which is defined as follows
\[
B_{ie}=\begin{cases}1, \qquad \mbox{if } e=(i,j)\\
-1, \qquad \mbox{if } e=(j,i)\\
0, \qquad \mbox{otherwise}, \end{cases}
\]
where an orientation is chosen arbitrarily for each edge.
Let $\widehat{W}\in\mathbb{R}^{M\times M}$ be a diagonal matrix where $\widehat{W}_{ee}=W_{ij}$ if $e=(i,j)$. Then, we can define the graph Laplacian $L$ as 
\begin{equation}
\label{laplacian}
L=B\widehat{W}B^T.
\end{equation}
It is important to underline that the graph Laplacian obtained using \eqref{laplacian} is independent from the edge orientation in $\G$.
\subsection{Graph Fourier Transform}
A graph signal $x\in\R^N$ in the vertex domain is a real-valued function defined on the nodes of the graph $\G$, such that $x_i$, $i=1,\dots,N$ is the value of the signal at node $i\in \V$ \cite{shuman2013emerging}. For example, for an image signal we can consider an associated graph where the nodes of the graph are the pixels of the image.
Then, the smoothness of $x$ on $\G$ can be measured using the Laplacian $L$ \cite{zhou2004regularization}
\begin{equation}
\label{smooth}
x^T L x=\frac{1}{2}\sum_{i=1}^N \sum_{j=1}^N W_{ij}(x_i-x_j)^2.
\end{equation}
Eq. \eqref{smooth} shows that a graph signal $x$ is considered to be smooth if strongly connected nodes have similar signal values. This equation also shows the importance of the graph. In fact, with a good graph representation the discontinuities should be penalized by low edge weights, in order to obtain a smooth representation of the signal. Finally, the eigenvectors of the Laplacian are used to define the  graph Fourier transform (GFT) \cite{shuman2013emerging} of the signal $x$ as follows: 
\[
\hat{x}=\Psi^T x.
\]
The graph signal $x$ can be easily retrieved from $\hat{x}$ by inversion, namely $x=\Psi \hat{x}$. Analogously to the Fourier transform in the Euclidean domain, the GFT is used to describe the graph signal in the Fourier domain. 

\subsection{Comparison between KLT and GFT}
As we have said in Section \ref{sec:rel_work}, the KLT is the transform that optimally decorrelates a signal that has Gaussian entries. In this section, we discuss the connection of the graph Fourier transform with the KLT, showing that the GFT can be seen as an approximation of the KLT. 

Let us consider a signal $x\in\mathbb{R}^N$ that follows a Gaussian Markov Random Field (GMRF) model with respect to a graph $\mathcal{G}$, with a mean $\mu$ and a precision matrix $Q$. Notice that the GMRF is a very generic model, where the precision matrix can be defined with much freedom, as long as its non-zero entries encode the partial correlations between random variables, and as long as their locations correspond to the edges of the graph. It has been proved that, if the precision matrix $Q$ of the GMRF model corresponds to the Laplacian $L$, then the KLT of the signal $x$ is equivalent to the GFT \cite{zhang2013analyzing}. 

As shown before, the graph Laplacian has a very specific structure where the non-zero components correspond to the edges of the graph, and, for this reason, it is a sparse matrix, since typically $|E|\ll N^2$. Since the precision matrix in general does not have such fixed structure, we now study the KLT of a signal whose model is a GMRF with a generic precision matrix $Q$. 
In this case, the GFT does not correspond to the KLT anymore and the GFT should be considered as an approximation of the KLT, where the precision matrix is forced to follow this specific structure. In order to find the GFT that best approximates the KLT, we introduce a maximum likelihood estimation problem, using an approach similar to the one presented in \cite{egilmez2016graph}. The density function of a GMRF has the following form \cite{rue2005gaussian}
\[
p(x)=(2\pi)^{-\frac{N}{2}}(\mbox{det }Q)^{\frac{1}{2}}\exp\left(-\frac{1}{2}(x-\mu)^T Q (x-\mu) \right).
\]
The log-likelihood function can then be computed as follows
\begin{equation}
\label{likelihood}
\log\mathcal{L}(Q,\mu|x)=\log (\mbox{det }Q)^{\frac{1}{2}}-\frac{1}{2}(x-\mu)^T Q(x-\mu).
\end{equation}
Given $x_1, ..., x_n$ observations of the signal $x$, we find the Laplacian matrix $L$ that best approximates $Q$ by solving the following problem
\begin{equation}
\label{prob_like}
\begin{split}
&\max_{L\in\Gamma}\  \log \mathcal{L}(L,\mu|x_1, ..., x_n),
\end{split}
\end{equation}
where  $\Gamma$ denotes the set of valid Laplacian matrices. Then, by using \eqref{likelihood}, the problem in \eqref{prob_like} can be written as
\begin{equation}
\label{eq:max_like}
\max_{L\in\Gamma}\  \log (\mbox{det}^*L)^{\frac{1}{2}}-\frac{1}{2}\mbox{tr}((X-\bar{\mu})^T L (X-\bar{\mu})),
\end{equation}
where $X$ is the matrix whose columns are the $N$ column vectors $x_1, x_2, ..., x_N$, and $\bar{\mu}$ is a matrix having $\mu$ as all its columns and $\mbox{det}^*$ is the pseudo-determinant (since $L$ is singular). 
The optimization problem in \eqref{eq:max_like} defines the graph whose GFT best approximates the KLT. The advantage of using the GFT instead of the KLT is that we force the precision matrix to follow the specific sparse structure defined by the Laplacian. In this way, the transform matrix can be transmitted to the decoder in a more compact way. However, even if the Laplacian matrix is more sparse than the precision matrix, the optimization problem in \eqref{eq:max_like} still does not explicitly take into account the cost of the graph transmision, which is very important in compression applications. For this reason, in the next section we present a new graph learning method that considers also this auxiliary cost and we highlight the connection between the proposed graph construction problem and the maximum likelihood estimation problem presented in \eqref{eq:max_like}.


\section{Graph transform optimization}
Graph-based compression methods use a graph representation of the signal through its GFT, in order to obtain a data-adaptive transform that captures the main characteristics of the signals. The GFT coefficients are then encoded, instead of the original signal values themselves. In general, if we consider a signal that is smooth with respect to its graph representation, its energy is concentrated in the low frequency coefficients of the GFT, hence it is easily compressible. In order to obtain good compression performance, the graph should therefore be chosen such that it leads to a smooth representation of the signal. At the same time, it should also be easy to encode, since it has to be transmitted to the decoder for signal reconstruction. This is a very important requirement, because often the cost of the graph representation outweighs the benefits of using an adaptive transform for signal representation. In order to find a good balance between graph signal representation benefits and its coding costs, we introduce a new graph construction approach that takes into consideration the above mentioned criteria.

We pose the problem of finding the optimal graph as a rate-distortion optimization problem defined as
\begin{equation}
\label{prob_orig}
\min_{L\in\mathbb{R}^{N\times N}} \mathcal{D}(L)+\gamma (\mathcal{R}_c(L) + \mathcal{R}_G(L)),
\end{equation}
 where $\mathcal{D}(L)$ is the distortion between the original signal and the reconstructed one. Instead, the total coding rate is composed of two representation costs, namely the cost of the signal transform coefficients $\mathcal{R}_c(L)$ and the cost of the graph description $\mathcal{R}_G(L)$. Each of these terms depends on the graph characterized by $L$ and on the coding scheme. We describe them in more details in the rest of the section.
 
\subsection{Distortion approximation}
We define the distortion $\mathcal{D}(L)$ as follows
\[
\mathcal{D}(L)=\lVert u-\tilde{u}(L)\rVert^2=\lVert \hat{u}(L)-\hat{u}_q(L)\rVert^2,
\] 
where $u$ and $\tilde{u}(L)$ are respectively the original and the reconstructed signal, and $\hat{u}(L)$ and $\hat{u}_q(L)$ are respectively the transform coefficients and the quantized transform coefficients. The equality holds due to the orthonormality of the GFT. If we consider a uniform scalar quantizer with the same step size $q$ for all the transform coefficients and if $q$ is small, the expected value of the distortion $\mathcal{D}(L)$ can be approximated as follows \cite{gray1998quantization}
\[
\mathcal{D}=\frac{q^2 N}{12}.
\]
With this high-resolution approximation, the distortion depends only on the quantization step size and it does not depend on the chosen $L$ \cite{hu2015multiresolution}. For simplicity, in the rest of the paper we adopt this assumption. Therefore, the optimization problem \eqref{prob_orig} is reduced to finding the graph that permits to minimize the rate terms.

\subsection{Rate approximation of the transform coefficients}

We can evaluate the cost of the transform coefficients $\mathcal{R}_c(L)$ by evaluating the smoothness of the signal on the graph described by $L$. We use the approximation proposed in \cite{hu2015multiresolution}, \cite{kim2012graph}, namely
\begin{equation}
\label{R_c}
\begin{split}
\mathcal{R}_c(L)&=u^T L u=u^T \left( \sum_{l=0}^{N-1}\lambda_l(L) \psi_l(L) \psi_l(L)^T\right)u\\
&=\sum_{l=0}^{N-1}\lambda_l(L)(u^T\psi_l(L))(\psi_l(L)^T u)=\sum_{l=0}^{N-1}\lambda_l \hat{u}_l^2(L),
\end{split}
\end{equation}
where $\lambda_l(L)$ and $\psi_l(L)$ are respectively the $l$-th eigenvalue and $l$-th eigenvector of $L$. Therefore, $\mathcal{R}_c(L)$ is an eigenvalue-weighted sum of squared transform coefficients. In this way, it assumes that the coding rate decreases when the smoothness of the signal over the graph defined by $L$ increases. In addition, \eqref{R_c} relates the measure of the signal smoothness with the sparsity of the transform coefficients. The approximation in \eqref{R_c} does not take into account the coefficients that corresponds to $\lambda_0(L)=0$ (i.e., the DC coefficients). Thus, \eqref{R_c} does not capture the variable cost of DC coefficients in cases where the graph contains a variable number of connected components. However,  in our work we ignore this cost as we impose that the graph is connected.

It is also interesting to point out that there is a strong connection between \eqref{R_c} and \eqref{eq:max_like}. In fact, if we suppose that $\mu=0$ and if we consider $u$ as the only observation of the signal $x$, then the second term of the log-likelihood in  \eqref{eq:max_like} is equal to $-\mathcal{R}_c(L)$. For this reason, we can say that the solution of our optimization problem can be seen as an approximation of the KLT.

\subsection{Rate approximation of the graph description}
\label{sec:graph_des}
The graph description cost $\mathcal{R}_G(L)$ depends on the method that is used to code the graph.  Generally, a graph could have an arbitrary topology. However, in order to reduce the graph transmission cost, we choose to use a fixed incidence matrix $B$ for the graph and to vary only the edge weights.
Therefore, the graph can be defined simply by a vector $w\in\mathbb{R}^M$, where $w_e$ with $1\le e\le M$ is the weight of the edge $e$. Then, by using \eqref{laplacian} we can define the graph Laplacian $L=B^T\mbox{diag}(w)B$.

\begin{figure}[tb]
\centering
\minipage{0.4\columnwidth}
\centering
  \includegraphics[width=\linewidth]{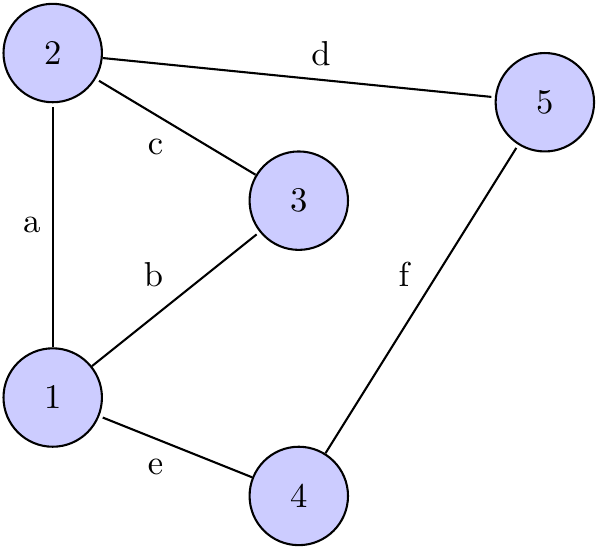}
  (a)
\endminipage\hspace{0.4cm}
\minipage{0.4\columnwidth}
\centering
  \includegraphics[width=\linewidth]{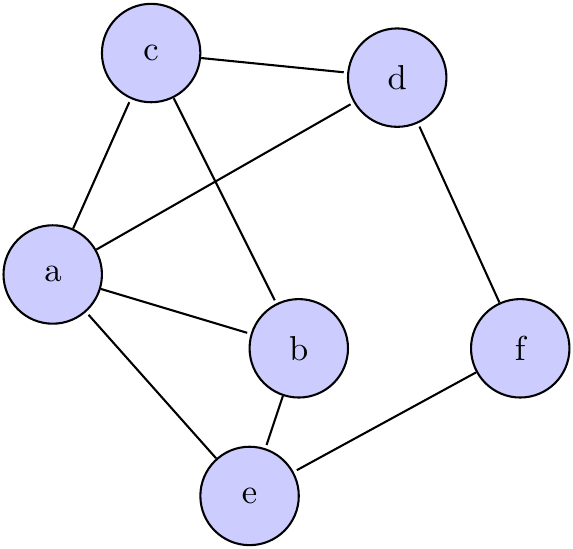}
  (b)
\endminipage
 \caption{An example of a graph (a) and its corresponding dual graph (b). The edges in the first graph (labeled with lower case letters) become the nodes of the corresponding dual graph.}
\label{fig:dualgraph}
\vspace{-0.3cm}
\end{figure}

In order to compress the edge weight vector $w$, we propose to treat it as a graph signal that lies on the dual graph $\G_d$. Given a graph $\G$, we define its dual graph $\G_d$ as an unweighted graph where each node of $\G_d$ represents an edge of $\G$ and two nodes of $\G_d$ are connected if and only if their corresponding edges in $\G$ share a common endpoint.  An example of a dual graph is shown in Fig. \ref{fig:dualgraph}. 
We choose to use this graph representation for the edge weight signal $w$ because consecutive edges  $\G$ often have similar weights, since the signals have often smooth regions or smooth transitions between regions. The latter is generally true in case of images. In this way,  the dual graph can provide a smooth representation of $w$.
We can define the graph Laplacian matrix $L_d\in\mathbb{R}^{M\times M}$ of the dual graph $\G_d$ and the corresponding eigenvector and eigenvalue matrices $\Psi_d\in\mathbb{R}^{M\times M}$ and $\Lambda_d\in\mathbb{R}^{M\times M}$ such that $L_d=\Psi_d \Lambda_d \Psi_d^T$. We highlight that, since $\G_d$ is an unweighted graph, it is independent of the choice of $L$, and by consequence also $\Lambda_d$ and $\Psi_d$ are independent from $L$.

Since $w$ can be represented as a graph signal, we can compute its GFT  $\hat{w}\in\mathbb{R}^M$ as
\[
\hat{w}=\Psi_d^T w.
\]
Therefore, we can use $\hat{w}$ to describe the graph $\G$ and we evaluate the cost of the graph description by measuring the coding cost of $\hat{w}$. It has been shown that the total bit budget needed to code a vector is proportional to the number of non-zero coefficients \cite{mallat1998analysis}, thus we approximate the cost of the graph description by measuring the sparsity of $\hat{w}$ as follows
\begin{equation}
\label{R_G}
\mathcal{R}_G(L)=\lVert \hat{w} \rVert_1=\lVert \Psi_d^T w \rVert_1.
\end{equation}
We highlight that we use two different types of approximations for $\mathcal{R}_c(L)$ and $\mathcal{R}_G(L)$, even if both of them are treated as graph signals. This is due to the fact that the two signals have different characteristics. In the case of an image signal $u$, we impose that the signal is smooth over $\G$, building the graph $\G$ with this purpose. Instead for $w$, even if we suppose that consecutive edges usually have similar values, we have no guarantees that $w$ is smooth on $\G_d$, since $\G_d$ is fixed and it is not adapted to the image signal. Therefore, in the second case using a sparsity constraint is more appropriate for capturing the characteristics of the edge weight signal $w$.

To be complete, we finally note that the dual graph has already been used in graph learning problems in the literature. In particular, in \cite{liu2015joint} the authors propose a method for joint denoising and contrast enhancement of images using the graph Laplacian operator, where the weights of the graph are defined through an optimization problem that involves the dual graph. Moreover, \cite{hu2016graph} presents a graph-based dequantization method by jointly optimizing the desired graph-signal and the similarity graph, where the weights of the graph are treated as another graph signal defined on the dual graph. The approximation of $\mathcal{R}_G(L)$ presented in \eqref{R_G} may look similar to the one used in \cite{hu2016graph}. The main difference between the two formulations is that  in \eqref{R_G} we minimize the sparsity of $w$ in the GFT domain in order to lossy code the signal $w$; instead, in \cite{hu2016graph}, the authors minimize the differences between neighboring edges in order to optimize the graph structure without actually coding it.

\subsection{Graph construction problem}

By using \eqref{laplacian}, \eqref{R_c} and \eqref{R_G}, our graph construction problem \eqref{prob_orig} is reduced to the following optimization problem
\begin{equation}
\label{gl-problem}
\min_{w\in\mathbb{R}^M} u^TB(\mbox{diag}(w))B^T u+\alpha \lVert \Psi^T_d w \rVert_1,
\end{equation}
where $\alpha$ is a weighting constant parameter, that allows us to balance the contribution of the two terms.

\begin{figure*}[tb]
\centering
  \includegraphics[width=17cm]{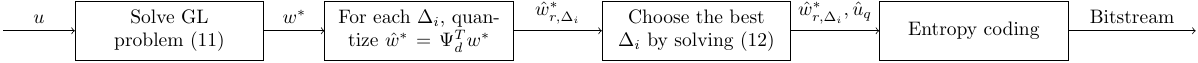}
 \caption{Block diagram of the proposed coding method for an input image $u$.}
\label{fig:block}
\end{figure*}

Building on the rate-distortion formulation of \eqref{gl-problem}, we design the graph by solving the following optimization problem
\begin{equation}
\label{obj_fun}
\begin{split}
&\min_{w\in\mathbb{R}^{M}} u^T B(\mbox{diag}(w))B^Tu+\alpha \lVert \Psi_d^T w \rVert_1-\beta \mathbf{1}^T\mbox{log}(w),\\
&\ \mbox{ s. t.}\quad w\le \mathbf{1},
\end{split}
\end{equation}
where $\alpha$ and $\beta$ are two positive regularization parameters and $\mathbf{1}$ denotes the constant one vector.  The inequality constraint has been added to guarantee that all the weights are in the range $(0,1]$, which is the same range of the most common normalized weighting functions \cite{grady2010discrete}. Then, the logarithmic term has been added to penalize low weight values and to avoid the trivial solution. In addition, this term guarantees that $w_m>0$, $\forall m$, so that the graph is always connected. A logarithmic barrier is often employed in graph learning problems \cite{kalofolias2016learn}. In particular, it has further been shown that a graph with Gaussian weights can be seen as the result of a graph learning problem with a specific logarithmic barrier on the edge weights\cite{kalofolias2016learn}. 

The problem in \eqref{obj_fun} can be cast as a convex optimization problem with a unique minimizer. To solve this problem, we write the first term in the following form
\[
\begin{split}
u^T B(\mbox{diag}(w))B^Tu&= \mbox{tr}((B^Tuu^T B)\mbox{diag}(w))\\
&=\mbox{vec}(B^Tuu^TB)^T\mbox{vec}(\mbox{diag}(w))\\
&=\mbox{vec}(B^Tuu^TB)^TM_{\mbox{\tiny diag}}w,
\end{split}
\]
where $\mbox{tr}(\cdot)$ denotes the trace of a matrix, $\mbox{vec}(\cdot)$ is the vectorization operator, and $M_{\mbox{\tiny diag}}\in\mathbb{R}^{M^2\times M}$ is a matrix that converts the vector $w$ into $\mbox{vec}(\mbox{diag}(w))$. Then, we can rewrite problem \eqref{obj_fun} as
\begin{equation}
\label{problem_final}
\begin{split}
&\min_{w\in\mathbb{R}^{M}} \mbox{vec}(B^Tuu^TB)^TM_{\mbox{\tiny diag}}w+\alpha \lVert \Psi_d^T w \rVert_1-\beta \mathbf{1}^T\mbox{log}(w),\\
&\ \mbox{ s. t.}\quad w\le \mathbf{1}.
\end{split}
\end{equation}
The problem in \eqref{problem_final} is a convex problem with respect to the variable $w$ and can be solved via interior-point methods \cite{boyd2004convex} whose complexity is approximately $O(M^3)$ (for the worst case).

\section{Graph-based image compression}
\label{sec:comp}
We now describe how the graph construction problem of the previous section can be applied to block-based image compression. 
It is important to underline that the main goal of this section is to present an application of our framework. Therefore, we do not present an optimization of the full coding process, but we mainly focus on the transform block. 

As pointed out in the previous sections, given an image block $u$ we have two different types of information to transmit to the decoder: the transform coefficients of the image signal $\hat{u}$ and the description of the graph $\hat{w}$. The image coefficients $\hat{u}$ are quantized and then coded using an entropy coder. Under the assumption of high bitrate, the optimal entropy-constrained quantizer is the uniform quantizer \cite{wiegand2010source}. Moreover, it has been proved  that, under the assumption that all the transform coefficients follow the same probability distribution, the transform code is optimized when the quantization steps of all coefficients are equal \cite{mallat2008wavelet}. For these reasons, we quantize the image transform coefficients $\hat{u}$ using a uniform quantizer with the same step size $q$ for all the coefficients. Then, since we assume that the non-zero coefficients are concentrated in the low frequencies, we code the quantized coefficients until the last non-zero coefficient using an adaptive bitplane arithmetic encoder \cite{witten1987arithmetic} and we transmit the position of the last significant coefficient. 

The graph itself is transmitted by its GFT coefficients vector $\hat{w}$, which is quantized and then transmitted to the decoder using an entropy coder. 
In order to reduce the cost of the graph description, we reduce the number of elements in $\hat{w}$ by taking into account only the first $\widetilde{M}\ll M$ coefficients, which usually are the most significant ones, and setting the other $M-\widetilde{M}$ coefficients to zero. The reduced signal $\hat{w}_r\in\mathbb{R}^{\widetilde{M}}$ is quantized using the same step size for all its coefficients and then coded with the same entropy coder used for the image signal.

Given an image signal, we first solve the optimization problem in \eqref{problem_final} obtaining the optimal solution $w^*$. To transmit $w^*$ to the decoder, we first compute its GFT coefficients $\hat{w}^*$ and the reduced vector $\hat{w}^*_r$, then we quantize $\hat{w}^*_r$ and code it using the entropy coder described above. It is important to underline that, since we perform a quantization of $\hat{w}^*_r$, the reconstructed signal $\tilde{w}^*$ is not strictly equal to the original $w^*$ and its quality depends on the quantization step size used. The graph described by $\tilde{w}^*$ is then used to define the GFT transform for the image signal.

\begin{figure*}[tb]
\minipage{0.29\textwidth}
  \includegraphics[width=\linewidth]{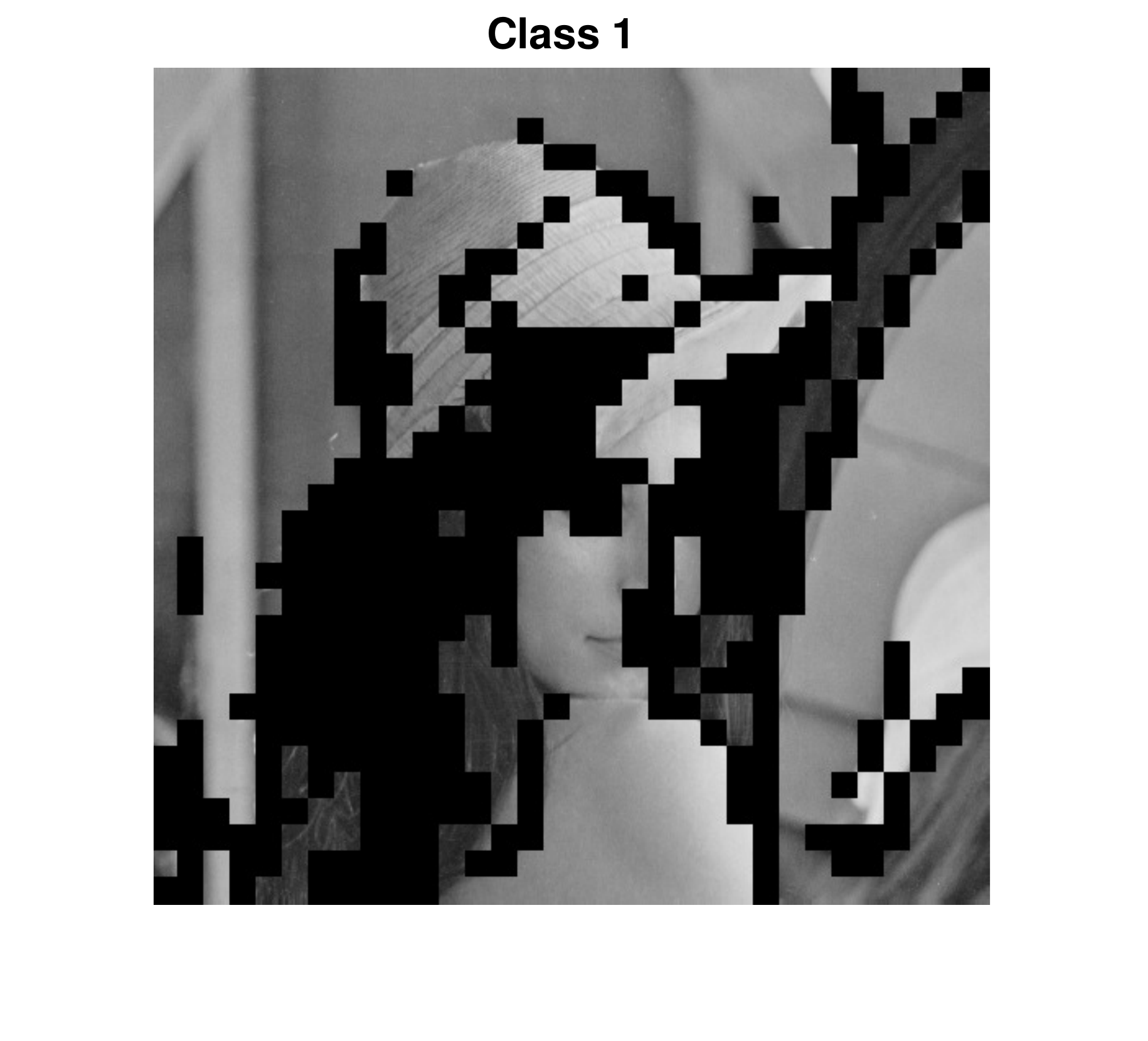}
\endminipage\hfill
\minipage{0.29\textwidth}
  \includegraphics[width=\linewidth]{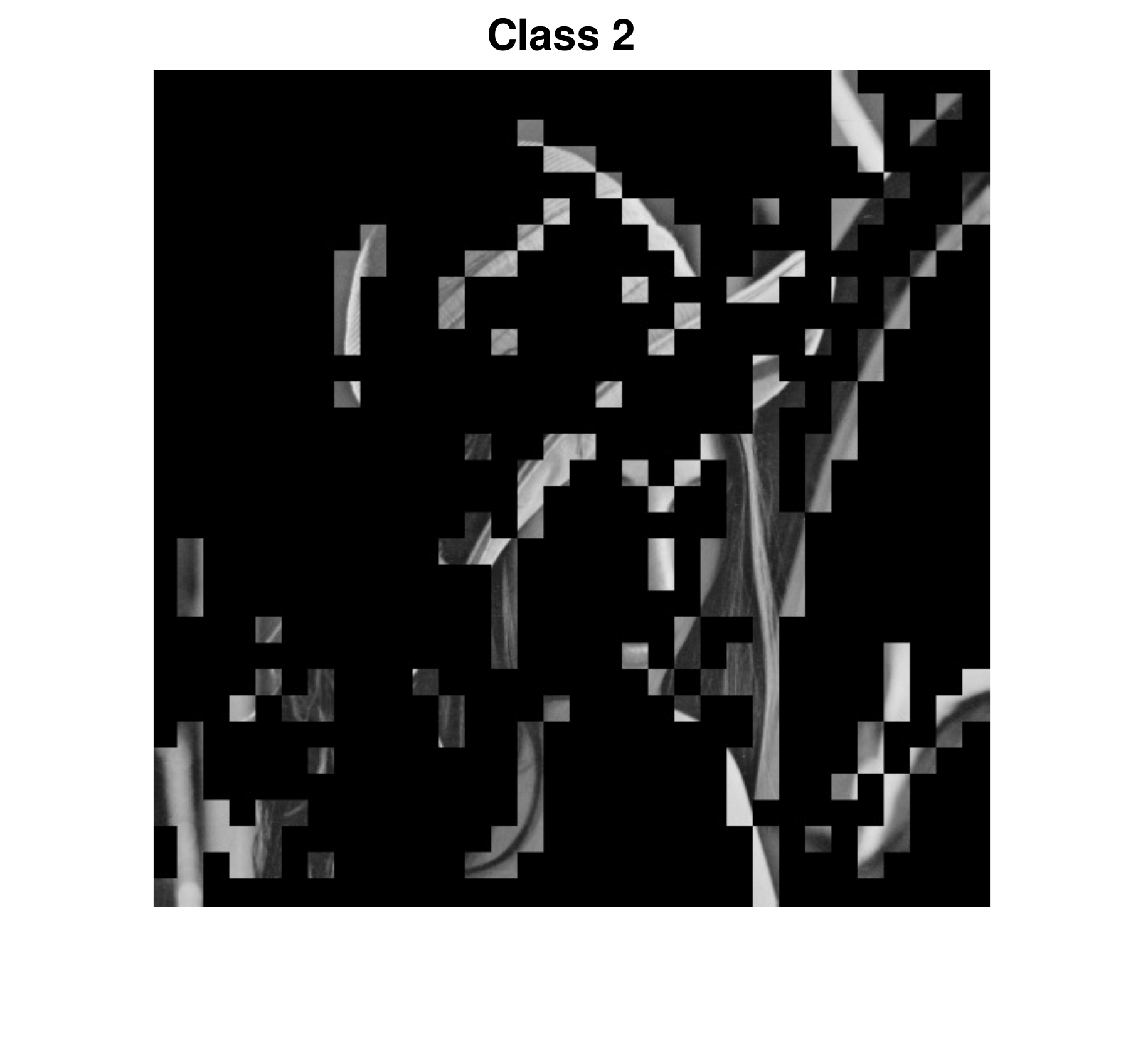}
\endminipage\hfill
\minipage{0.29\textwidth}
  \includegraphics[width=\linewidth]{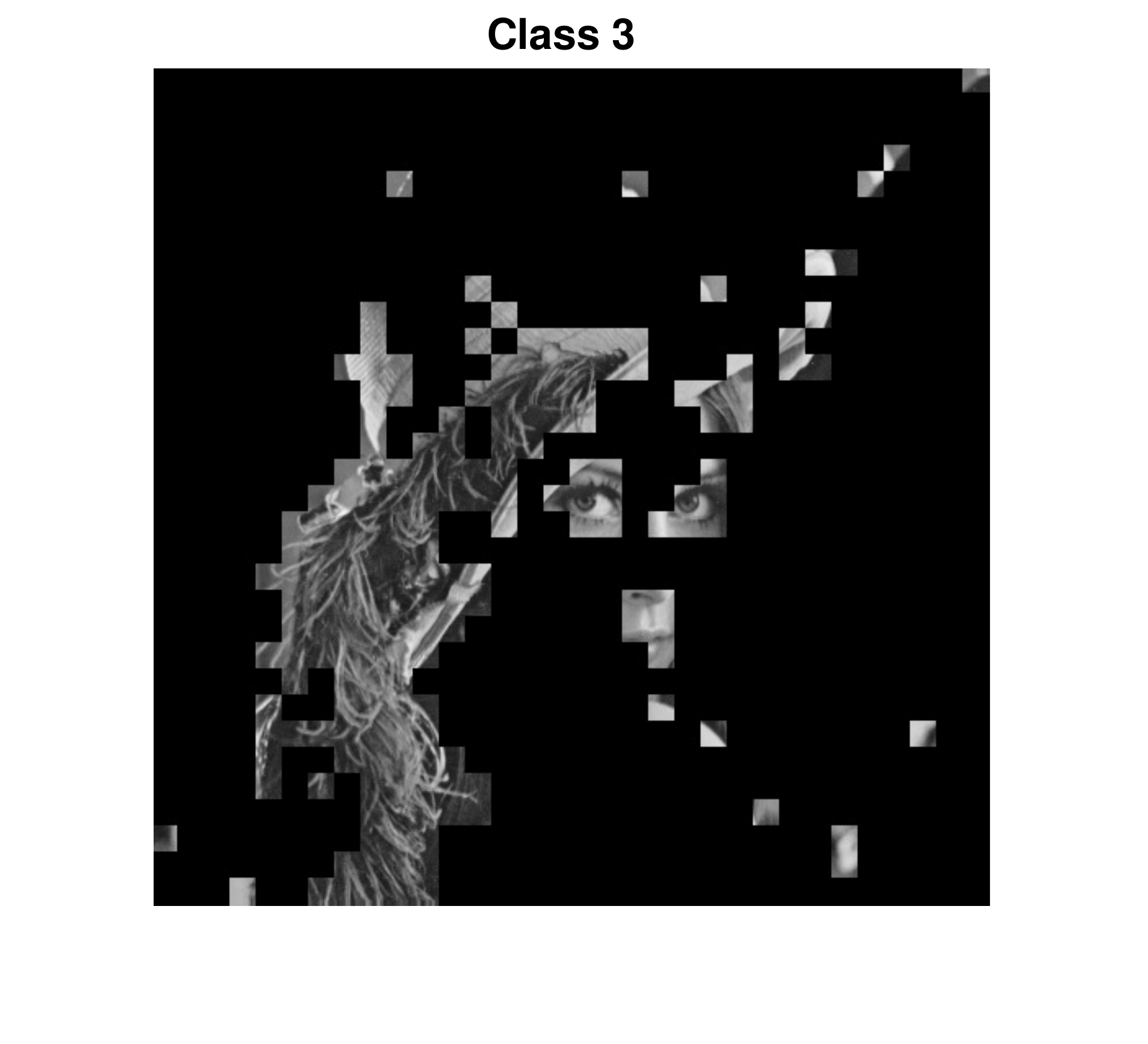}
\endminipage
\vspace{-0.2cm}
 \caption{Block classification of the image Lena. Class 1 contains smooth blocks, class 2 blocks with a dominant principal gradient, and  
class 3 consists of blocks with a more complex structure.}
\label{fig:block_class}
\end{figure*}

Since it is important to find the best tradeoff between the quality of the graph and its transmission cost, for each block in an image we test different quantization step sizes $\{\Delta_i\}_{1\le i\le Q}$ for a given graph represented by $\hat{w}^*_r$. To choose the best quantization step size, we use the following rate-distortion problem
\begin{equation} 
\label{prob:quant}
\min_{i} \mathcal{D}(\Delta_i)+\gamma(\mathcal{R}_c(\Delta_i)+\mathcal{R}_G(\Delta_i)),
\end{equation}
where $\mathcal{R}_G(\Delta_i)$ is the rate of $\hat{w}_{r,\Delta_i}^*$, the coefficient vector $\hat{w}_r^*$ quantized with $\Delta_i$, and $\mathcal{D}(\Delta_i)$ and $\mathcal{R}_c(\Delta_i)$ are respectively the distortion and the rate of the reconstructed image signal obtained using the graph transform described by  $\hat{w}_{r,\Delta_i}^*$. We point out that the choice of $\Delta_i$ depends on the quantization step size $q$ used for the image transform coefficients $\hat{u}$. In fact, at high bitrate (small $q$) we expect to have a smaller $\Delta_i$ and thus a more precise graph, instead at low bitrate (large $q$) we will have a larger $\Delta_i$ that corresponds to a coarser graph approximation. We also underline that, in \eqref{prob:quant}, we evaluate the actual distortion and rate without using the approximation introduced previously in \eqref{prob_orig}, \eqref{R_c}, \eqref{R_G}.  The actual coding methods described above are used to compute the rates $\mathcal{R}_c(\Delta_i)$ and $\mathcal{R}_G(\Delta_i)$. The principal steps of the proposed image compression method are summarized in Fig. \ref{fig:block}.

\subsection{Computational complexity}
The proposed image compression method requires to compute, both at the encoder and at the decoder, the eigenvectors of the Laplacian matrix.  The computational cost of this operation is approximately $O(N^3)$, where we recall that $N$ is the number of pixels in a block. Moreover, at the encoder it is also necessary to solve the optimization problem in \eqref{problem_final}, whose complexity is at most $O(M^3)$, where $M$ is the number of edges and, since we use a grid graph, $M$ is lower than $2N$. For this reason, the overall complexity of the proposed image compression method is higher at the encoder than at the decoder. This could be an advantage in asymmetric applications, where the encoder may afford more complexity than the decoder. In Table \ref{tab:ex_time} we show the execution times of the proposed method for encoding and decoding an image block. We also perform a comparison with other three techniques, namely the method proposed in \cite{hu2015multiresolution}, a baseline DCT coder and a graph-based method where the graph is designed using a Gaussian weight function (see Section VI and VII for more details on these methods). The execution time of the proposed method at the encoder is higher than the other three methods, this is probaby due to the fact that for the optimization problem we use a general purpose solver which is not optimized for our specific problem. Instead, at the decoder we can see that the performance of the proposed method are  comparable with the ones of \cite{hu2015multiresolution} and the Gaussian method.
\begin{table}[t]
\centering
\caption{Execution time computed using an Intel Core i7 @ 4.7 GHz processor.}
\renewcommand{\arraystretch}{1.1}
\begin{tabular}{|c|c|c|}

\hline
Method      &Encoder&Decoder\\
\hline
Proposed       &$2.36$ s&$4.10 \cdot 10^{-3}$ s\\
\hline
DCT      &$6.95 \cdot10^{-4}$ s&$6.15 \cdot10^{-4}$ s\\
\hline
\cite{hu2015multiresolution}          &$3.55 \cdot10^{-2}$ s&$1.10 \cdot10^{-3}$ s\\
\hline
Gaussian        &$3.57 \cdot10^{-2}$ s&$4.30 \cdot10^{-3}$ s\\
\hline
\end{tabular}
\label{tab:ex_time}
\end{table}
\section{Experimental results on natural images}
In this section, we evaluate the performance of our illustrative graph-based encoder for natural images. We first describe the general experimental settings, then we present the obtained experimental results.

\subsection{Experimental setup}
First of all, we subdivide the image into non-overlapping 16$\times$16 pixel blocks. For each block, we define the edge weights using the graph learning problem described in the previous sections.
The chosen topology of the graph is a 4-connected grid: this is the most common graph topology for graph-based image compression, since its number of edges is not too high, and thus the coding cost is limited. In a 4-connected square grid with $N$ nodes, we have $M=2\sqrt{N}(\sqrt{N}-1)$ edges.  In all our experiments on natural images, we use $Q=8$ possible quantization step sizes $\Delta_i$ for $\hat{w}_r$ and we set $\widetilde{M}=64$, which is the length of the reduced coefficient vector $\hat{w}_r$. In order to set the value of the parameter $\alpha$ in \eqref{problem_final}, we first have to perform a block classification. In fact, we recall that the parameter $\alpha$ in \eqref{problem_final} is related to the $l_1$-norm of $\hat{w}$, where $\hat{w}$ are the GFT coefficients of the signal $w$ that lies on the dual graph. As we have explained previously, the motivation for using the dual graph is that consecutive edges usually have similar values. However, this statement is not always true, but it depends on the characteristics of the block. In smooth blocks nearly all the edges will have similar values. Instead, in piecewise smooth blocks there could be a small percentage of edges whose consecutive ones have significantly different values. Finally, in textured blocks this percentage may even increase in a significant way. For this reason, we perform a priori a block classification using a structure tensor analysis, as done in \cite{rotondo2015designing}. The structure tensor is a matrix derived from the gradient of an image patch, and it is commonly used in many image processing algorithms, such as edge detection \cite{kothe2003edge}, corner detection \cite{harris1988combined,kenney2005axiomatic} and feature extraction \cite{forstner1986feature}. Let $\mu_1$ and $\mu_2$ be the two eigenvalues of the structure tensor, where $\mu_1\ge\mu_2\ge 0$. We classify the image blocks in the following way:
\begin{itemize}
\item Class 1: smooth blocks, if $\mu_1\approx\mu_2\approx 0$;
\item Class 2: blocks with a dominant principal gradient, if $\mu_1\gg\mu_2\approx 0$; 
\item Class 3: blocks with a more complex structure, if $\mu_1$ and $\mu_2$ are both large. 
\end{itemize}
Fig. \ref{fig:block_class}  shows an example of block classification.
For each block class, we have set the values of parameters $\alpha$ and $\beta$  by fine tuning. We set $\alpha=100$ for blocks that belong to the first class, $\alpha=500$ for blocks that belong to the second class and $\alpha=800$ for blocks that belong to the third class. For all the three classes, we set the same value for the other optimization parameter, i.e., $\beta=1$.

\begin{table*}
\centering
\caption{Bjontegaard average gain in PSNR for natural images w.r.t. DCT}
\renewcommand{\arraystretch}{1.1}
\begin{tabular}{c|c|c|c||c|c|c||c|c|c||c|c|c|}
\cline{2-13}
      &\multicolumn{3}{c||} {Class 1} & \multicolumn{3}{c||}{Class 2} &\multicolumn{3}{c||}{Class 3}&\multicolumn{3}{c|}{Total}\\
\hline
\multicolumn{1}{|c|}{Image}&\begin{tabular}{@{}c@{}}Learned \\ graph\end{tabular}&\begin{tabular}{@{}c@{}}Gaussian \\ graph\end{tabular}&KLT&\begin{tabular}{@{}c@{}}Learned \\ graph\end{tabular}&\begin{tabular}{@{}c@{}}Gaussian \\ graph\end{tabular}&KLT&\begin{tabular}{@{}c@{}}Learned \\ graph\end{tabular}&\begin{tabular}{@{}c@{}}Gaussian \\ graph\end{tabular}&KLT&\begin{tabular}{@{}c@{}}Learned \\ graph\end{tabular}&\begin{tabular}{@{}c@{}}Gaussian \\ graph\end{tabular}&KLT\\
\hline
\multicolumn{1}{|c|}{Lena}        &0.11&0.11&0.15&0.70&0.49&0.03&0.51&0.32&0.10&0.31&0.23&0.13\\
\hline
\multicolumn{1}{|c|}{Boat}        &0.09& 0.07&0.21&0.33&0.21&0.11&0.46&0.34&0.07&0.28&0.21&0.16\\
\hline
\multicolumn{1}{|c|}{Peppers}   &0.18 &0.14&0.44&0.87 &0.45&0.25&0.88 &0.57&0.10&0.47&0.31&0.37\\
\hline
\multicolumn{1}{|c|}{House}     &0.05&0.05&0.07&0.68&0.59&0.11&0.61&0.49&0.01&0.36&0.31&0.08\\
\hline
\multicolumn{1}{|c|}{Couple}     &0.26&0.29&0.07&1.17&0.75&0.04&0.98&0.81&0.04&0.66&0.55&0.06\\
\hline
\multicolumn{1}{|c|}{Stream}     &0.17&0.19&0.18&0.50&0.30&0.01&0.31&0.22&0.13&0.31&0.23&0.13\\
\hline
\multicolumn{1}{|c|}{Barbara}     &0.13&0.16&0.18&0.37&0.28&0.14&0.30&0.22&0.02&0.26&0.22&0.12\\
\hline
\multicolumn{1}{|c|}{F16}     &0.12&0.20&0.10&0.79&0.70&0.13&0.79&0.66&0.05&0.48&0.46&0.10\\
\hline
\multicolumn{1}{|c|}{Female}     &0.14&0.15&0.12&1.22&0.84&0.06&1.14&0.88&0.07&0.55&0.46&0.11\\
\hline
\multicolumn{1}{|c|}{Lake}     &0.12&0.19&0.50&0.25&0.08&0.28&0.60&0.46&0.17&0.43&0.37&0.32\\
\hline
\end{tabular}
\label{tab:bjon}
\end{table*}

We compare the performance of the proposed method to a baseline coding scheme built on the classical DCT transform. In order to obtain comparable results, we code the transform coefficients $\hat{u}$ of the image signal using the same entropy coder for the graph-based method and for the DCT-based encoder. In the first case, in addition to the bitrate of $\hat{u}$, we count the bitrate due to the transmission of $\hat{w}^*_{i}$  and $\log Q$ additional bits per block to transmit the chosen quantization step size $\Delta_i$ for $\hat{w}_r$. We point out that we do not need to send any information about the block class, because this classification is only needed at the encoder in order to define the parameters of the optimization problem \eqref{problem_final}. For both methods, we vary the quantization step size $q$ of the transform coefficients to vary the encoding rates.   In addition, in our method, for each block, we compare the RD-cost of the GFT and the one of the DCT. Then, we eventually code the block with the transform that has the lowest RD-cost and we use 1 additional bit per block to signal if we are using the GFT or the DCT.

In order to show the advantages of the proposed graph construction problem, we compare our method with a classical graph construction technique that uses a Gaussian weight function \cite{grady2010discrete} to define the edge weights
\[
W_{ij}=e^{-\frac{(u_i-u_j)^2}{\sigma^2}},
\]
where $\sigma$ is a gaussian parameter that we defined as $\sigma=0.15\max_{i,j}|u_i-u_j|$. In order to have comparable results, we use the coding scheme described in Sec. V also for the Gaussian graph. 

We compare the proposed method also with an approximation of the KLT. We subdivide the image blocks into classes using the same block classification proposed above and for each class we compute a separate KLT. In this way, the only auxiliary information that we have to send to the decoder is the class index. In order to have a fair comparison,  the transform coefficients are coded with the same coding scheme that we have used for the other methods.

\begin{figure*}[tb]
\begin{minipage}[b]{0.5\linewidth}
  \centering
  \centerline{\includegraphics[width=7cm]{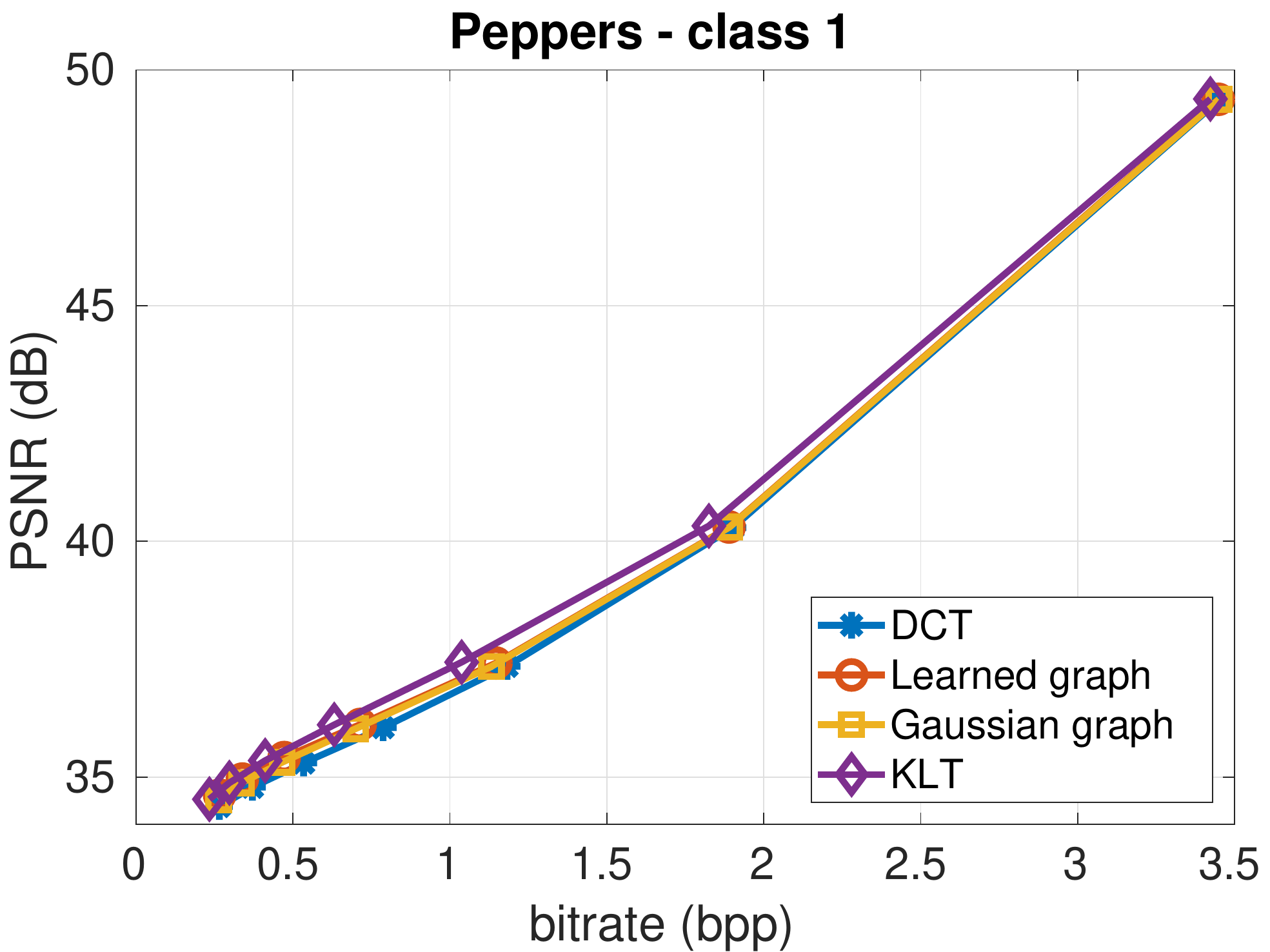}}
   \end{minipage}
\hfill
\begin{minipage}[b]{0.5\linewidth}
  \centering
  \centerline{\includegraphics[width=7cm]{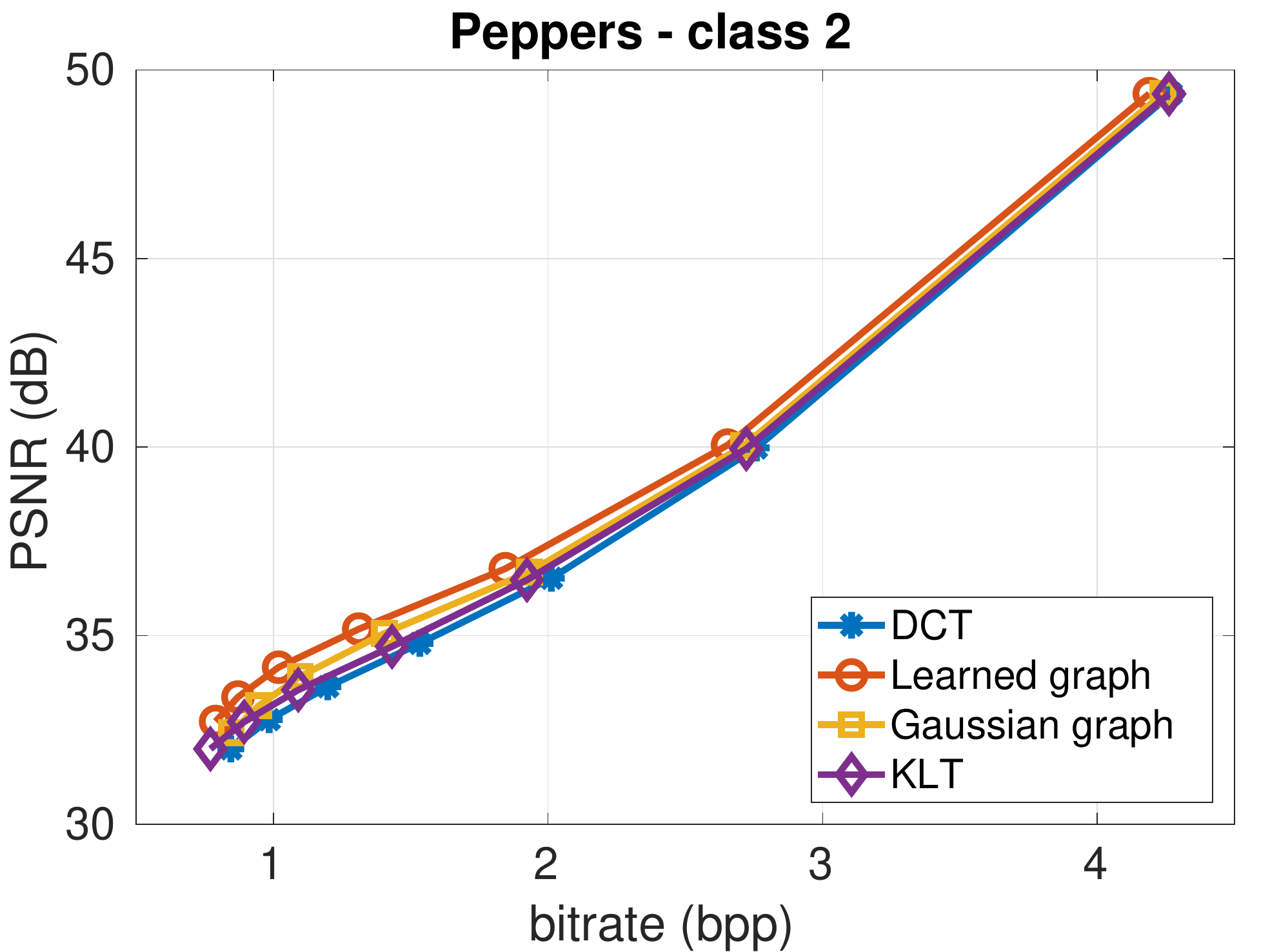}}
   \end{minipage}
\vfill
\begin{minipage}[b]{0.5\linewidth}
  \centering
  \centerline{\includegraphics[width=7cm]{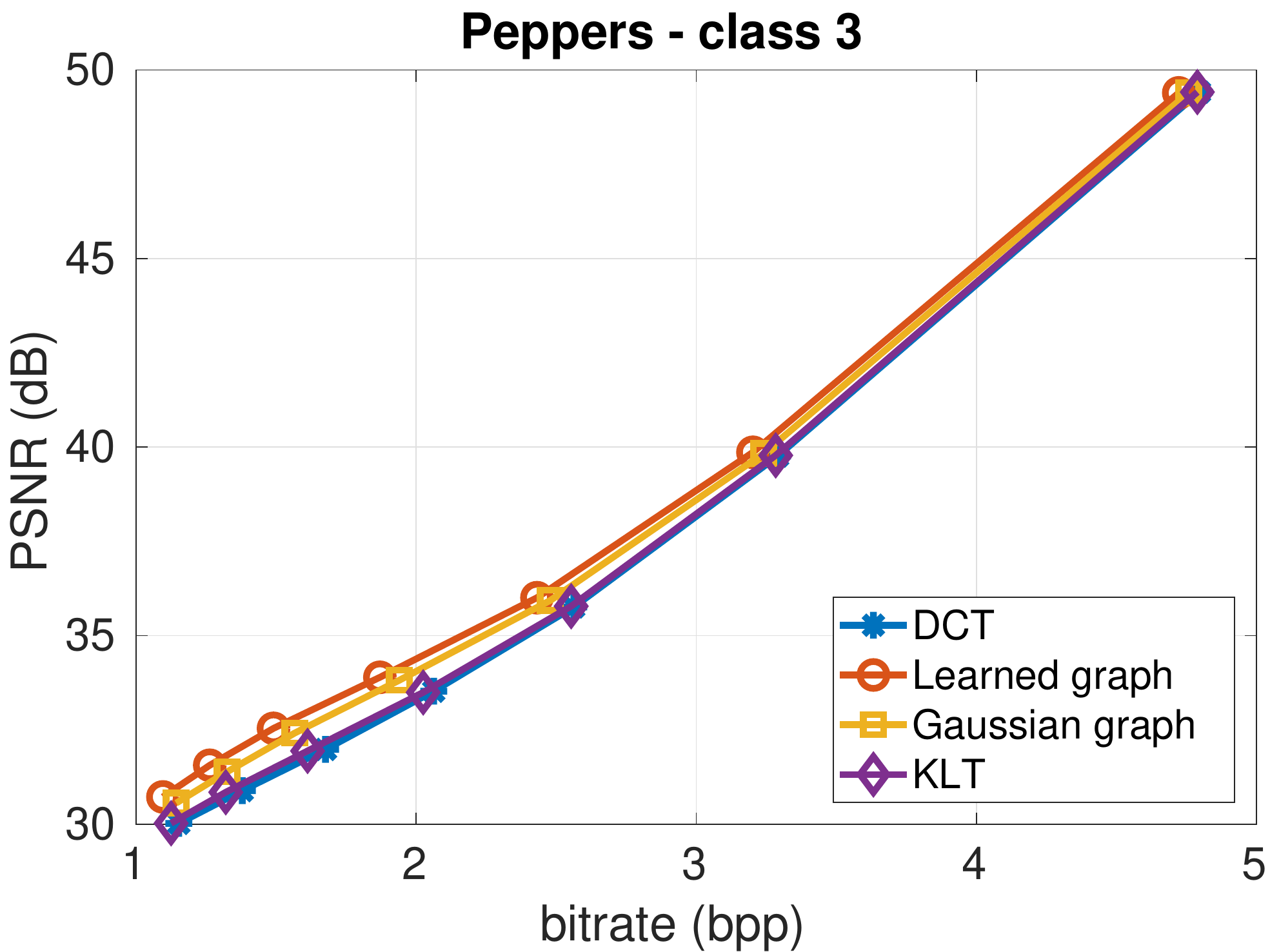}}
  \end{minipage}
\hfill
\begin{minipage}[b]{0.5\linewidth}
  \centering
  \centerline{\includegraphics[width=7cm]{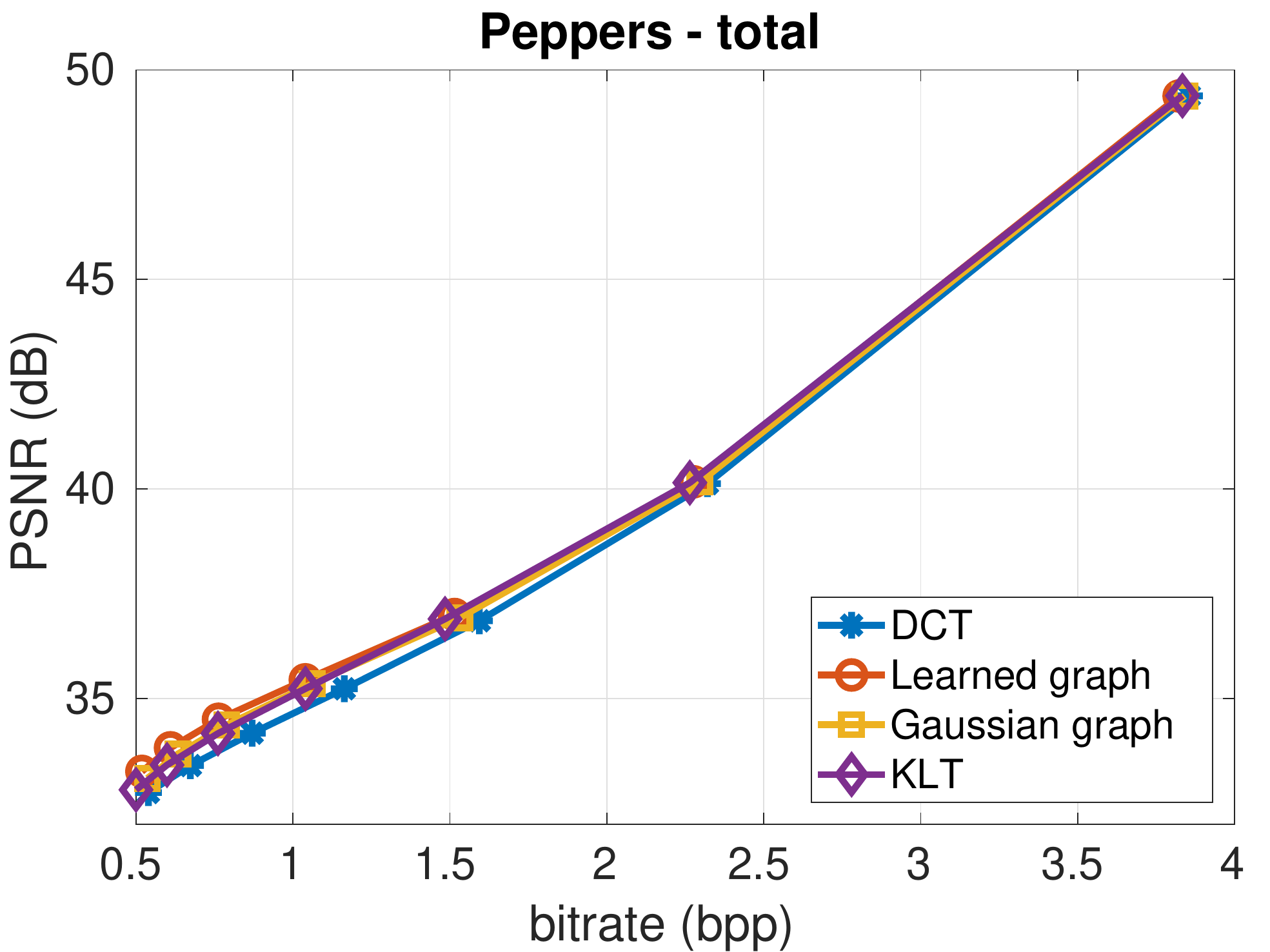}}
  \end{minipage}
 \caption{Rate-distortion curves for the image Peppers.}
\label{fig:RDc}
\end{figure*}

\subsection{Results}
\begin{figure}
\begin{minipage}[b]{0.3\linewidth}
\centering
\centerline{\includegraphics[width=\linewidth]{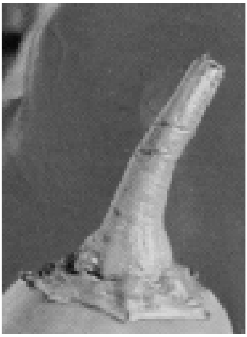}}
Original image
\end{minipage}
\hfill
\begin{minipage}[b]{0.3\linewidth}
\centering
\centerline{\includegraphics[width=\linewidth]{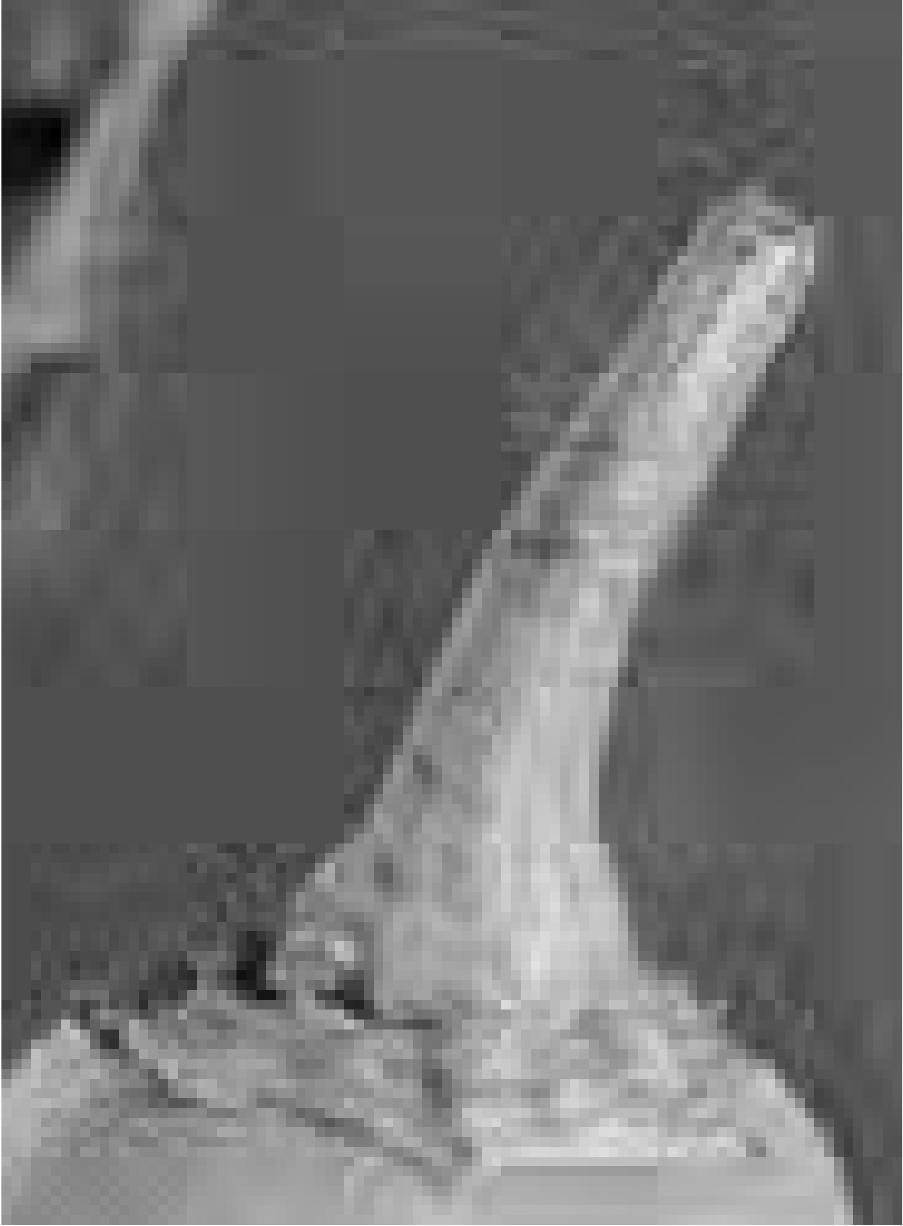}}
DCT
\end{minipage}
\hfill
\begin{minipage}[b]{0.3\linewidth}
\centering
\centerline{\includegraphics[width=\linewidth]{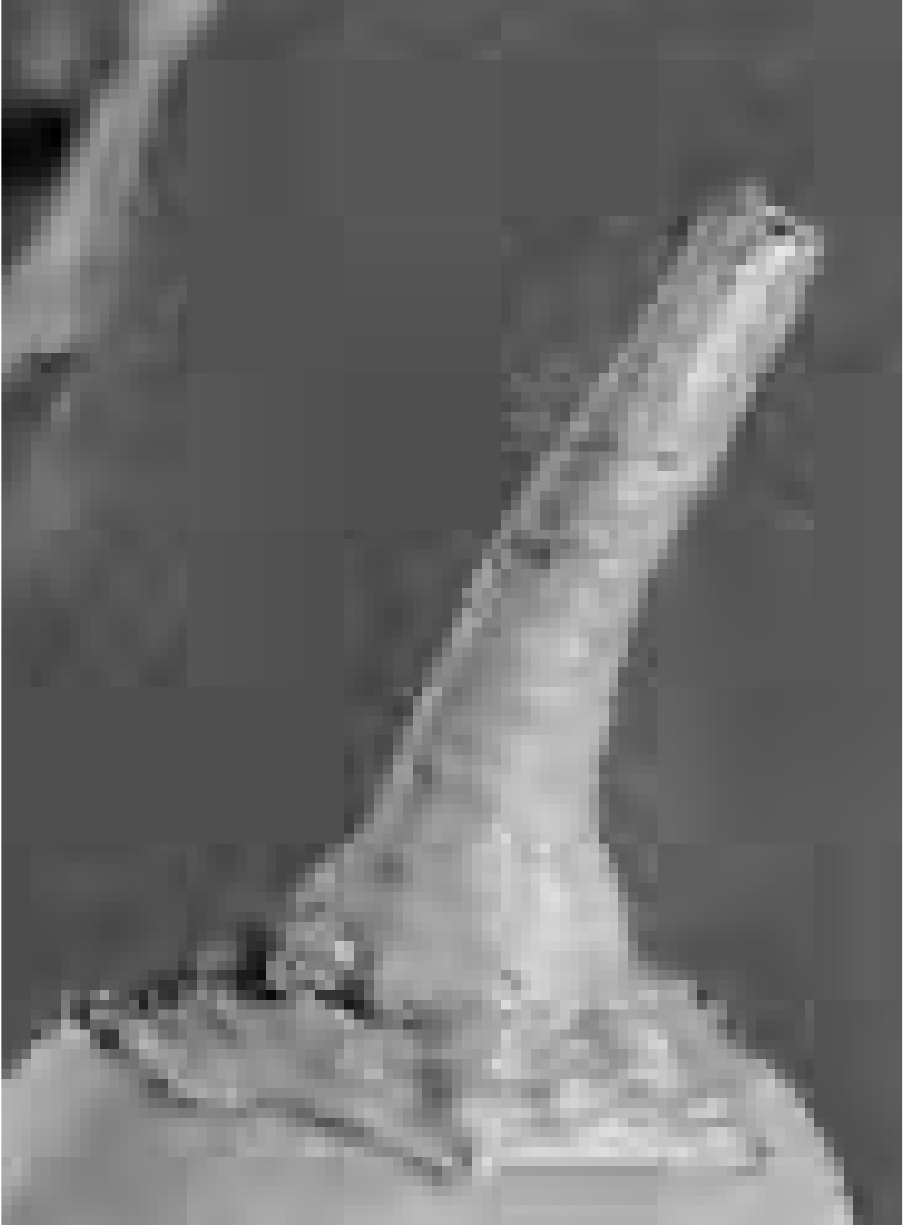}}
Proposed method
\end{minipage}
\caption{Visual comparison for a detail of the image Peppers at 0.6 bpp.}
\label{fig:natim_vis}
\end{figure}

The experiments are performed on ten classical grayscale images (House, Lena, Boat, Peppers, Stream, Couple, Stream, Barbara, F16, Female, Lake) \cite{img_db}. This dataset contains different types of natural images, for example some of them have smooth regions (e.g. House and Peppers), others instead are more textured (e.g. Boat, Lena, Couple and Stream). In Table \ref{tab:bjon}, we show the obtained performance results in terms of average gain in PSNR compared to DCT, evaluated through the Bjontegaard metric \cite{bjontegaard2001calcuation}.  Moreover, in Fig. \ref{fig:RDc} we show the rate-distortion curves for the image Peppers. To visually understand the performance difference, in Fig. \ref{fig:natim_vis} we show a visual comparison between the DCT and the proposed method for the image Peppers.
We see that, in the second and third classes, the proposed method outperforms DCT providing an average PSNR gain of 0.69 dB for blocks in the second class and 0.66 dB for blocks in the third class. It should be pointed out that there is not a significant difference in performance between the second class and the third one. This probably is due to the fact that the proposed graph construction method is able to adapt the graph and its description cost to the characteristics of each block. Instead, in the first class, which corresponds to smooth blocks, the gain is nearly 0, as DCT in this case is already optimal. 

We also notice that the method with the Gaussian graph always outperforms the DCT. However, in the classes where the DCT is not optimal, the learned graph always provides better performance. This shows that the proposed graph compression method in conjunction with a standard graph construction technique can still outperform the DCT, even if the graph is not optimized. On the other hand, these results show also that we can reach a higher level of performance if we build the graph using the proposed graph learning method, which is targeted for image compression and takes into account the coding cost of both the image signal and the graph structure. This highlights the importance of the proposed optimization problem.

Moreover, we can see that in the second and third class the proposed graph-based method outperforms the KLT; instead in the first class in some cases the KLT provides better perfomance. This could be explained by the fact that we are using an approximation of the KLT that is the same for all the blocks that belong to a class. In fact, in some cases it could be important to have a transform that reflects the peculiar characteristics of the block that we are processing, especially for blocks with sharp discontinuities or textured blocks.

\begin{figure}[tb]
\vspace{-0.3cm}
\centering
\includegraphics[width=9.5cm]{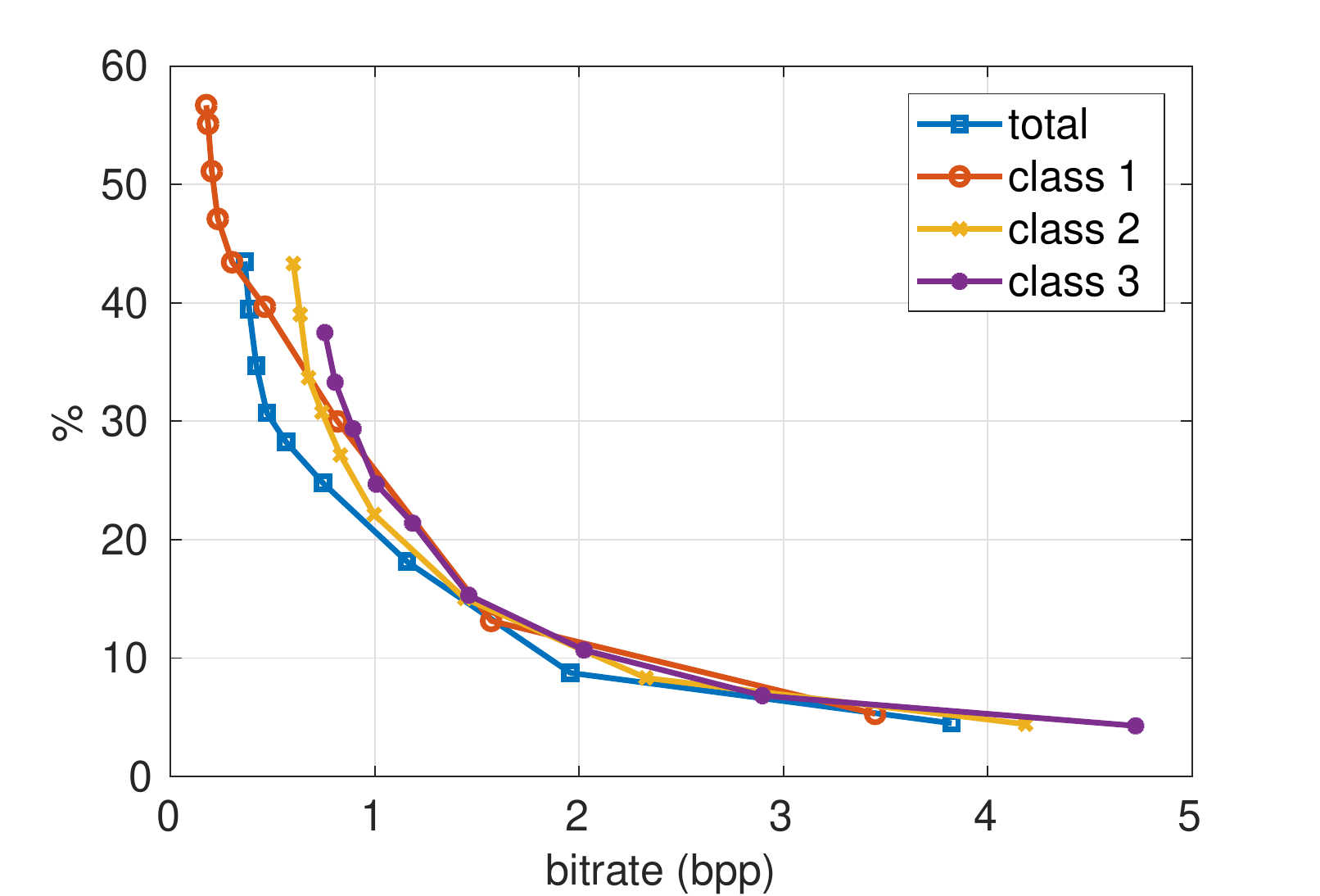}
\caption{Percentage of bitrate dedicated to graph transmission in function of the overall bitrate for the image Peppers.}
\label{fig:perc}
\end{figure}

Finally, we show in Figure \ref{fig:perc} the percentage of bitrate dedicated to the graph transmission in function of the overall bitrate. We can clearly see that the percentage of bitrate dedicated to the graph description increases as the bitrate decreases and the increase is steeper at very low bitrate. 

\subsection{Wavelet comparison}
For natural image compression, coding scheme based on wavelets have achieved significantly better performance compared to DCT-based compression methods \cite{taubman2012jpeg2000}. Differently from the DCT, wavelets are usually applied on the whole image without subdividing the image into blocks. In this way, it is possible to better exploit the spatial correlation of the pixels. Since in this paper we consider a block-based approach, a comparison with wavelets would not provide a fair assessment of the transform. For this reason, we decided to use as main benchmark the DCT. However, we also present a comparison between the proposed method and wavelets computed on a few sample images. In this case, we consider wavelets defined on blocks of different sizes. Table \ref{tab:wav} shows the obtained results. We evaluate the performance of the proposed method compared to wavelets. As we can see from the results, when the dimension of the block is small, the proposed method show a significant quality gain. Instead, at larger block size wavelets often outperform the proposed method. 

\begin{table}[t]
\centering
\caption{Bjontegaard average gain in PSNR compared to wavelets.}
\renewcommand{\arraystretch}{1.1}
\begin{tabular}{c|c|c|c|c|}
\cline{2-5}
      &\multicolumn{4}{c|} {Block size} \\
\hline
\multicolumn{1}{|c|}{Image}        &16$\times$16 &32$\times$32 & 64$\times$64 &128$\times$128\\
\hline
\multicolumn{1}{|c|}{Peppers}       &1.98&1.13&0.67&0.08\\
\hline
\multicolumn{1}{|c|}{Boat }       &1.46&0.34&-0.23&-0.44\\
\hline
\multicolumn{1}{|c|}{Stream}   &0.84&-0.30&-0.83&-0.98\\
\hline
  \multicolumn{1}{|c|}{Lake }   &1.45&0.72&0.35&0.15\\
\hline
\end{tabular}
\label{tab:wav}
\end{table}

\subsection{Validation of theoretical assumptions}
\label{sec:res_val}
\begin{figure}[t]
\vspace{-0.3cm}
\centering
\includegraphics[width=9cm]{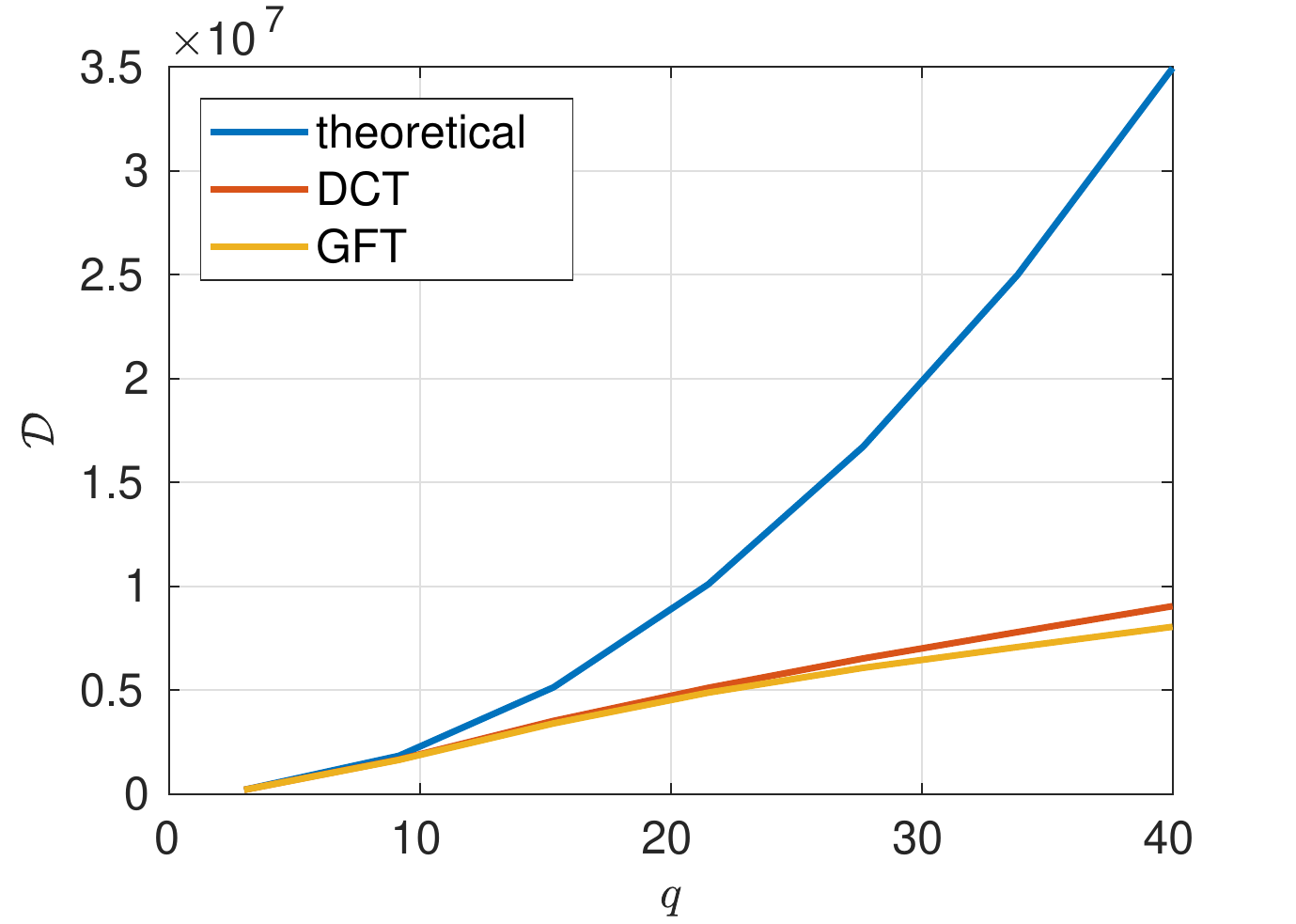}
\caption{Theoretical vs actual distortion with respect to the quantization step size $q$ for the image Peppers. The corresponding DCT and GFT bitrates lie approximately between 0.5 bpp (for $q=40$) and 3.8 bpp (for $q=3$) for both transforms. }
\label{fig:teo1}
\end{figure}

In this section, we examine to what extent the assumptions used to derive the graph optimization problem \eqref{problem_final} are validated.  The first assumption that we made is the approximation of the distortion presented in Sec. IV.A. In this case, we assume that the distortion depends only on the quantization step size. Figure \ref{fig:teo1} shows the theoretical distortion with respect to the quantization step size $q$ compared to the actual one. We can clearly see that, with small quantization step sizes (i.e. high bitrate), the theoretical distortion converges to the actual one. This is an expected result since this approximation is valid only at high bitrate. Then, we also examine the rate approximation presented in Sec. IV.B and IV.C. In this case, we compare the theoretical total bitrate computed using approximation \eqref{R_c} and \eqref{R_G} with the actual one. Table \ref{tab:corr} shows the correlation coefficient of these two quantities for a few sample images. For all the considered images, the correlation coefficient is close to one, showing that the rate approximation provides a good estimate of the actual bitrate.

\begin{table}[t]
\centering
\caption{Correlation coefficient between theoretical and actual bitrates.}
\renewcommand{\arraystretch}{1.1}
\begin{tabular}{|c|c|c|c|c|c|}
\hline
Image        &Peppers &Lena & Boat &Stream&Lake\\
\hline
Correlation      &0.954&0.987&0.977&0.998&0.988\\
\hline
\end{tabular}
\label{tab:corr}
\end{table}
\section{Experimental results on piecewise smooth images}
In this section, we evaluate the performance of the proposed method on piecewise smooth images, comparing our method with classical DCT and the state-of-the-art graph-based coding method of \cite{hu2015multiresolution}. We first describe the specific experimental setting used for this type of signals, then we present the obtained results.


\subsection{Experimental setup}
\begin{figure}[tb]
\begin{minipage}[b]{0.47\linewidth}
\centering
\centerline{\includegraphics[width=\linewidth]{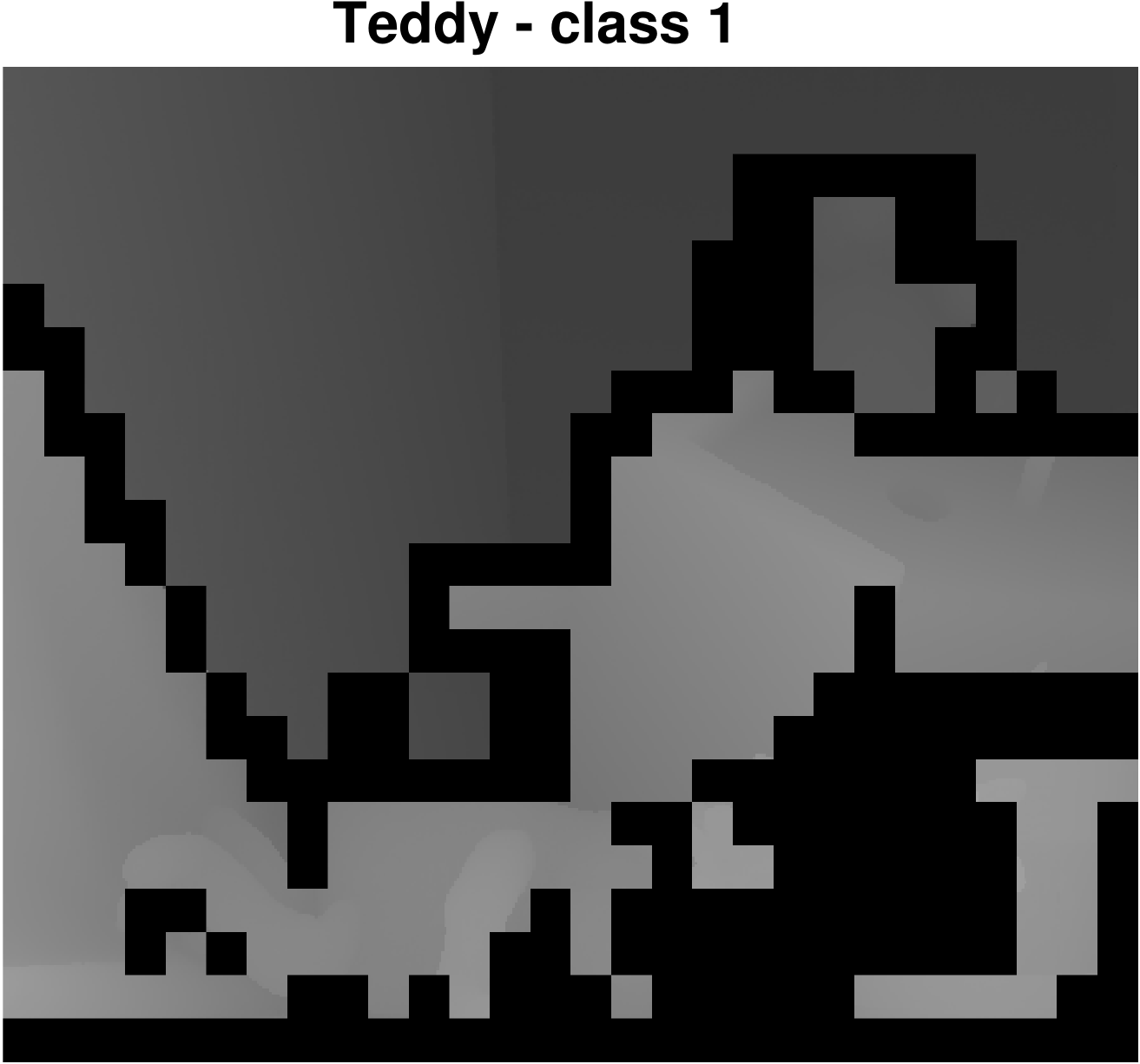}}
\end{minipage}
\hfill
\begin{minipage}[b]{0.47\linewidth}
\centering
\centerline{\includegraphics[width=\linewidth]{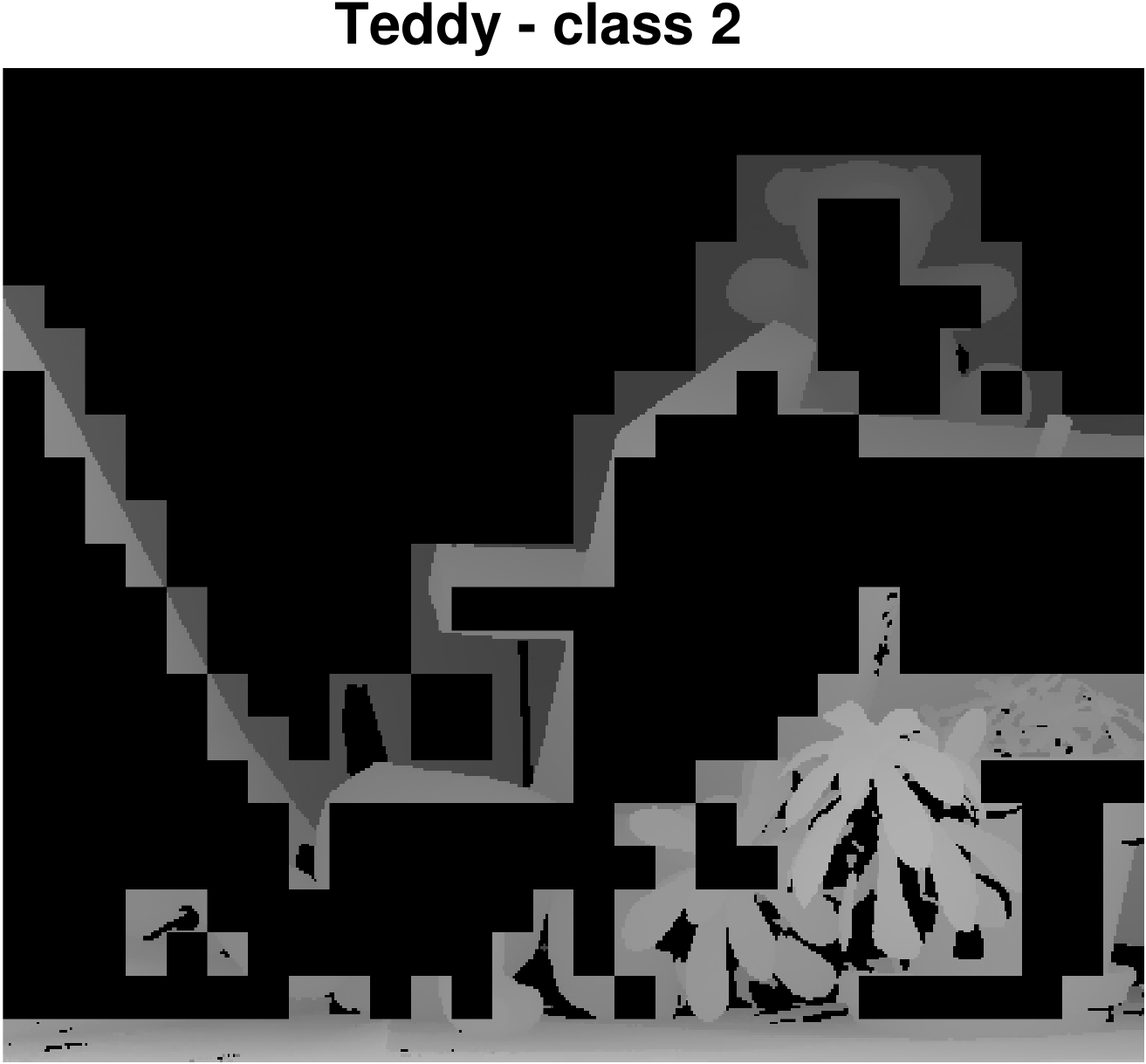}}
\end{minipage}
 \caption{Block classification of the image Teddy.}
\label{fig:block_class_pws}
\end{figure}

We choose as piecewise smooth signals ten depth maps taken from \cite{scharstein2003high,scharstein2007learning}. Similarly to the case of natural images, we split them into non-overlapping 16$\times$16 pixel blocks and the chosen graph topology is a 4-connected grid. In addition, we keep for $Q$ the same setting as the one used for natural images. 
Then, to define the parameters $\alpha$ and $\beta$ we again subdivide the image blocks into classes using the structure tensor analysis. In \cite{hu2015multiresolution}, the authors have identified three block classes for piecewise smooth images: smooth blocks, blocks with weak boundaries (e.g., boundaries between different parts of the same foreground/background) and blocks with sharp boundaries (e.g., boundaries between foreground and backgound). In our experiments, since we have observed that the first two classes have a similar behavior, we decided to consider only two different classes: 
\begin{itemize}
\item Class 1: smooth blocks and blocks with weak edges, if $\mu_1\approx\mu_2\approx 0$.
\item Class 2: blocks with sharp edges, if $\mu_1\gg 0$,
\end{itemize}
where $\mu_1$ and $\mu_2$, with $\mu_1\ge\mu_2\ge0$, are the two eigenvalues of the structure tensor. An example of block classification is shown in Fig. \ref{fig:block_class_pws}.
As done for natural images, for each class we set parameters $\alpha$ and $\beta$ by fine tuning. For the first class, we set $\alpha=40$ and $\beta=0.02$. For the second class, we set $\alpha=400$ and $\beta=1$.

\begin{figure}[tb]
\minipage{\linewidth}
\centering
  \includegraphics[width=0.6\linewidth]{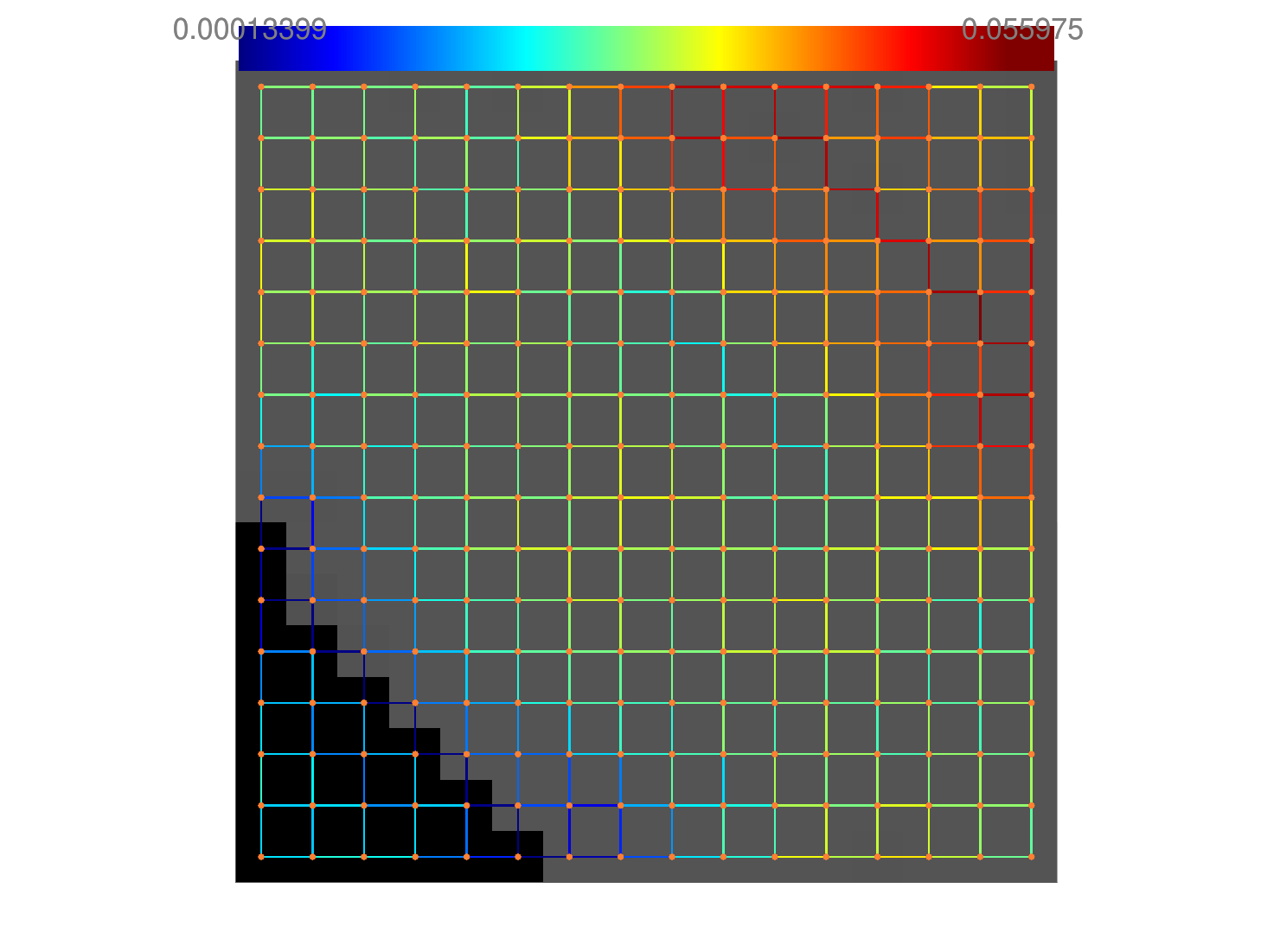}
\endminipage\vfill
\minipage{\linewidth}
\centering
  \includegraphics[width=0.65\linewidth]{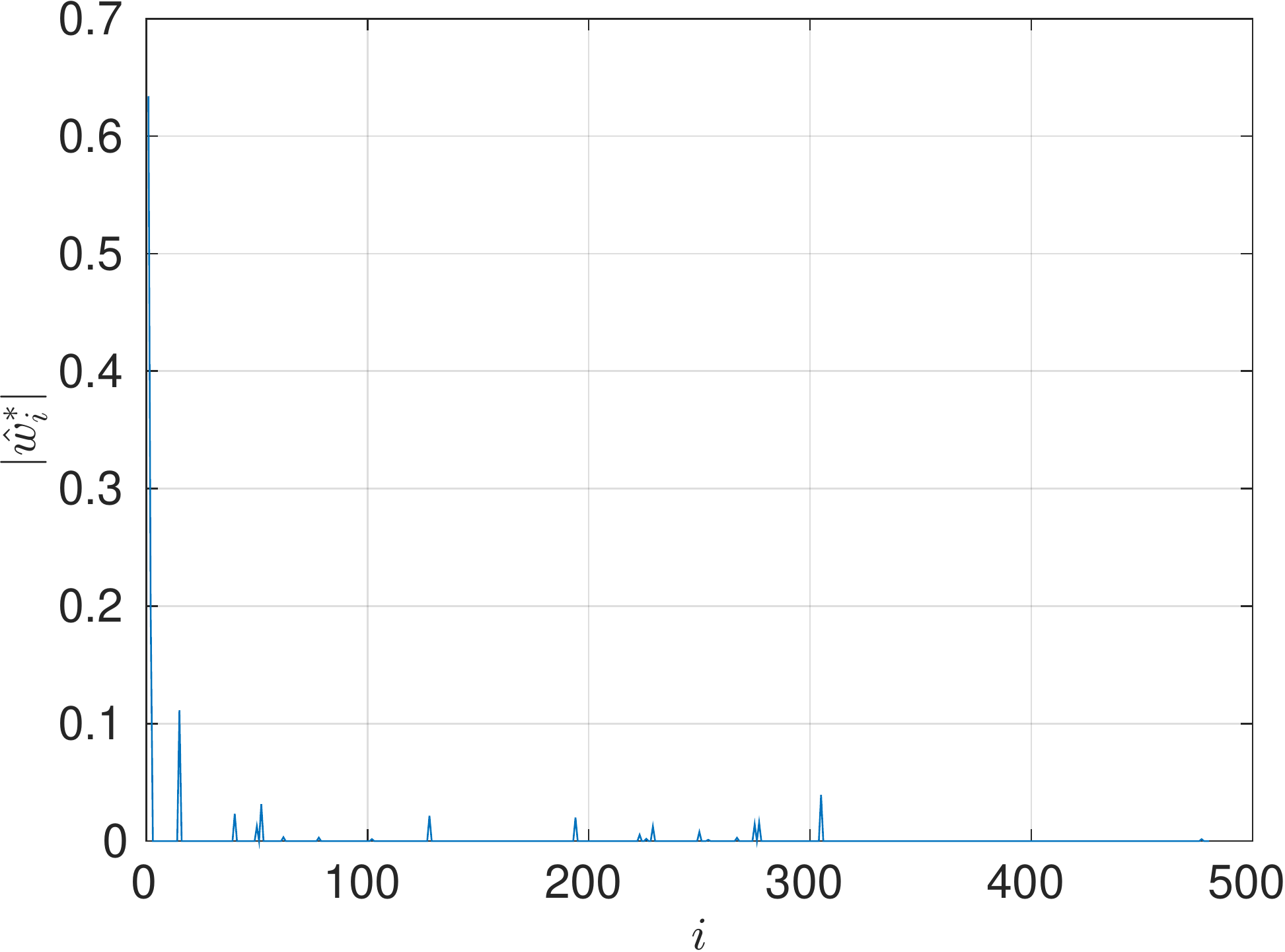}
\endminipage
 \caption{Top: Example of a piecewise smooth block and the corresponding learned graph. Bottom: The corresponding GFT coefficients $\hat{w}$.}
\label{fig:graph_example}
\end{figure}

With this type of signals, we have observed that the coefficients $\hat{w}^*$ of the learned graph are very sparse, as shown in Fig. \ref{fig:graph_example}. For this reason, we decided to modify the coding method used for $\hat{w}^*$. As done for natural images, we reduce the number of elements in $\hat{w}$ by taking into account only the first  $\widetilde{M}$ coefficients (in this case we set $\widetilde{M}=256$). Then, we use an adaptive binary arithmetic encoder to transmit a significance map that signals the non-zero coefficients. In this way, we can use an adaptive bitplane arithmetic encoder to code only the values of the non-zero coefficients. This allows a strong reduction of the number of coefficients that we have to transmit to the decoder. 

Similarly to the case of natural images, we compare our method to a transform coding method based on the classical DCT. However, in the specific case of depth map coding it has been shown that graph-based methods significantly outperform the classical DCT. 
For this reason, we also propose a comparison with a graph-based coding scheme that is specifically designed for piecewise smooth images. 
The method presented in \cite{hu2015multiresolution} achieves the state-of-the-art performance in graph-based depth image coding. This method uses a table-lookup based graph transform: the most popular GFTs are stored in a lookup table, and then for each block an exhaustive search is performed to choose the best GFT in rate-distortion terms. In this way, the side information that has to be sent to the decoder is only the table index. Moreover, the method in \cite{hu2015multiresolution} incorporates a number of coding tools, including multiresolution coding scheme and edge-aware intra-prediction. Since in our case we are interested in evaluating the performance of the transform, we only focus on the transform part and we use as reference method a simplified version of the method in \cite{hu2015multiresolution} that is similar to the one used in \cite{zhang2017graph}. The simplified version of \cite{hu2015multiresolution} that we implemented employs 16$\times$16 blocks and it does not make use of edge-aware prediction and multiresolution coding. Since the transform used in \cite{hu2015multiresolution} is based on a lookup table, we use 40 training depth images to build the table as suggested in \cite{hu2015multiresolution}. In the training phase, we identify the most common graph transforms. As a result, the obtained lookup table contains 718 transforms. Then, in the coding phase each block is coded using one of the transforms contained in the lookup table or the DCT. The coding method used for the table index is the same as in \cite{hu2015multiresolution}. Instead for the transform coefficients $\hat{u}$, in order to have comparable results, we use the coding method described in Sec. \ref{sec:comp}.

\begin{table}[t]
\centering
\caption{Bjontegaard average gain in PSNR compared to DCT.}
\renewcommand{\arraystretch}{1.1}
\begin{tabular}{|c|c|c|c|}
\hline
Image      &Class 1&Class 2&Total\\
\hline
Teddy       &0.59&6.12&4.82\\
\hline
Cones      &0.70&8.37&6.88\\
\hline
Art           &0.56&8.66&7.62\\
\hline
Dolls        &0.55&8.59&5.57\\
\hline
Moebius   &0.82&7.36&5.52\\
\hline
Reindeer   &0.45&8.34&5.75\\
\hline
Rocks   &0.36&11.41&7.05\\
\hline
Cloth   &0.48&11.91&7.22\\
\hline
Baby   &0.20&10.18&8.03\\
\hline
Aloe   &0.29&9.16&7.75\\
\hline
\end{tabular}
\label{tab:bjon_pws}
\end{table}

\begin{table}[t]
\centering
\caption{Bjontegaard average gain in PSNR between the proposed method and the reference method.}
\renewcommand{\arraystretch}{1.1}
\begin{tabular}{|c|c|c|c|}
\hline
      Image& Class 1 & Class 2& Total\\
\hline
Teddy  &-0.87&-0.12&-0.38\\
\hline
Cones   &-1.21&1.17&0.54\\
\hline
Art  &-0.89&0.86&0.49\\
\hline
Dolls   &-0.78&1.16&0.26\\
\hline
Moebius &-1.14&-0.47&-0.76\\
\hline
Reindeer &-0.51&0.02&-0.33\\
\hline
Rocks &-0.81&1.64&0.39\\
\hline
Cloth &-0.89&2.26&0.65\\
\hline
Baby   &-0.66&0.04&-0.24\\
\hline
Aloe   &-1.40&1.73&0.95\\
\hline
\end{tabular}
\label{tab:bjon_pws_comp}
\end{table}

\subsection{Results}

\begin{figure}[tb]
\begin{minipage}[b]{\linewidth}
  \centering
  \centerline{\includegraphics[width=7cm]{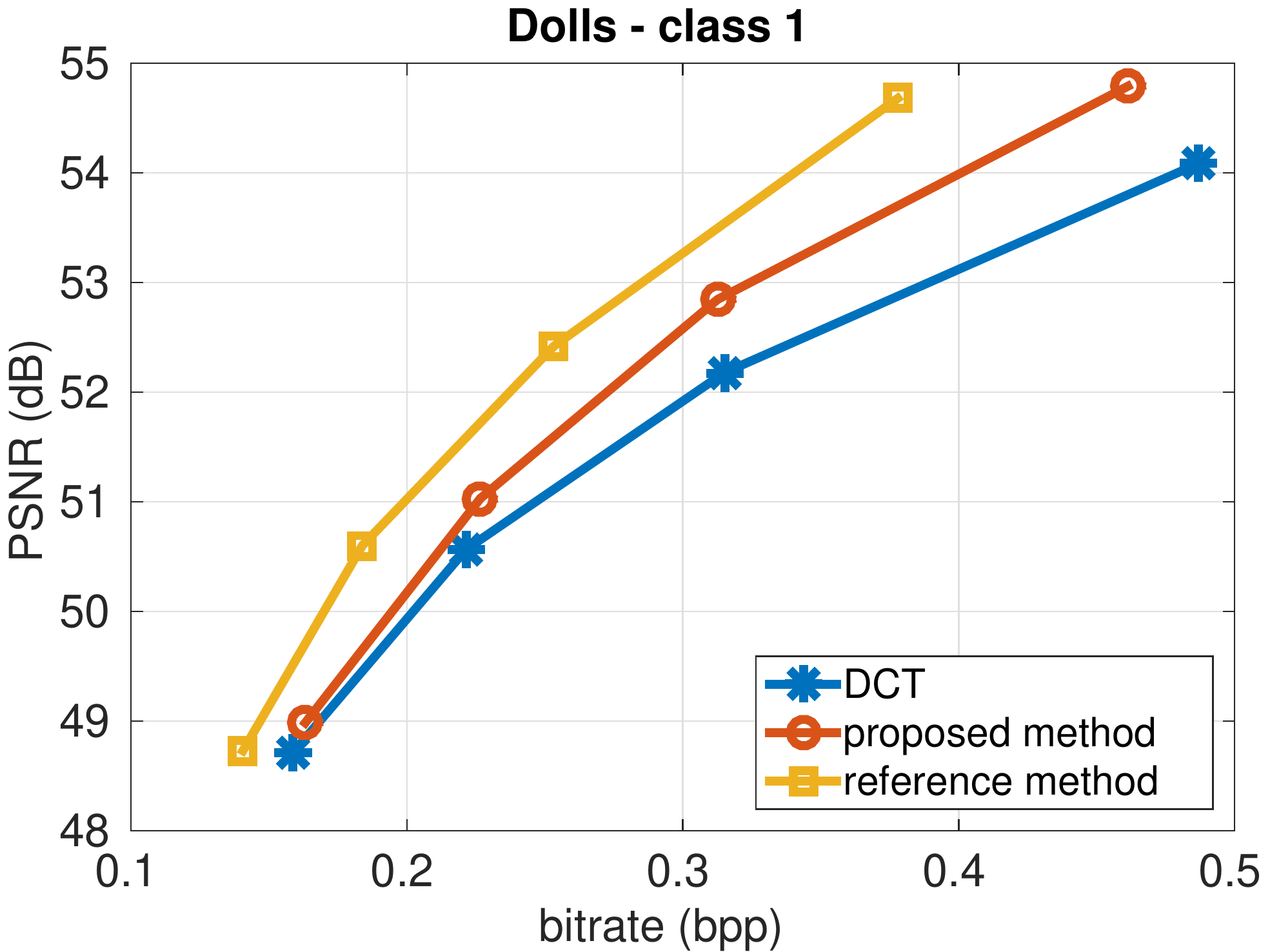}}
   \end{minipage}
\vfill
\begin{minipage}[b]{\linewidth}
  \centering
  \centerline{\includegraphics[width=7cm]{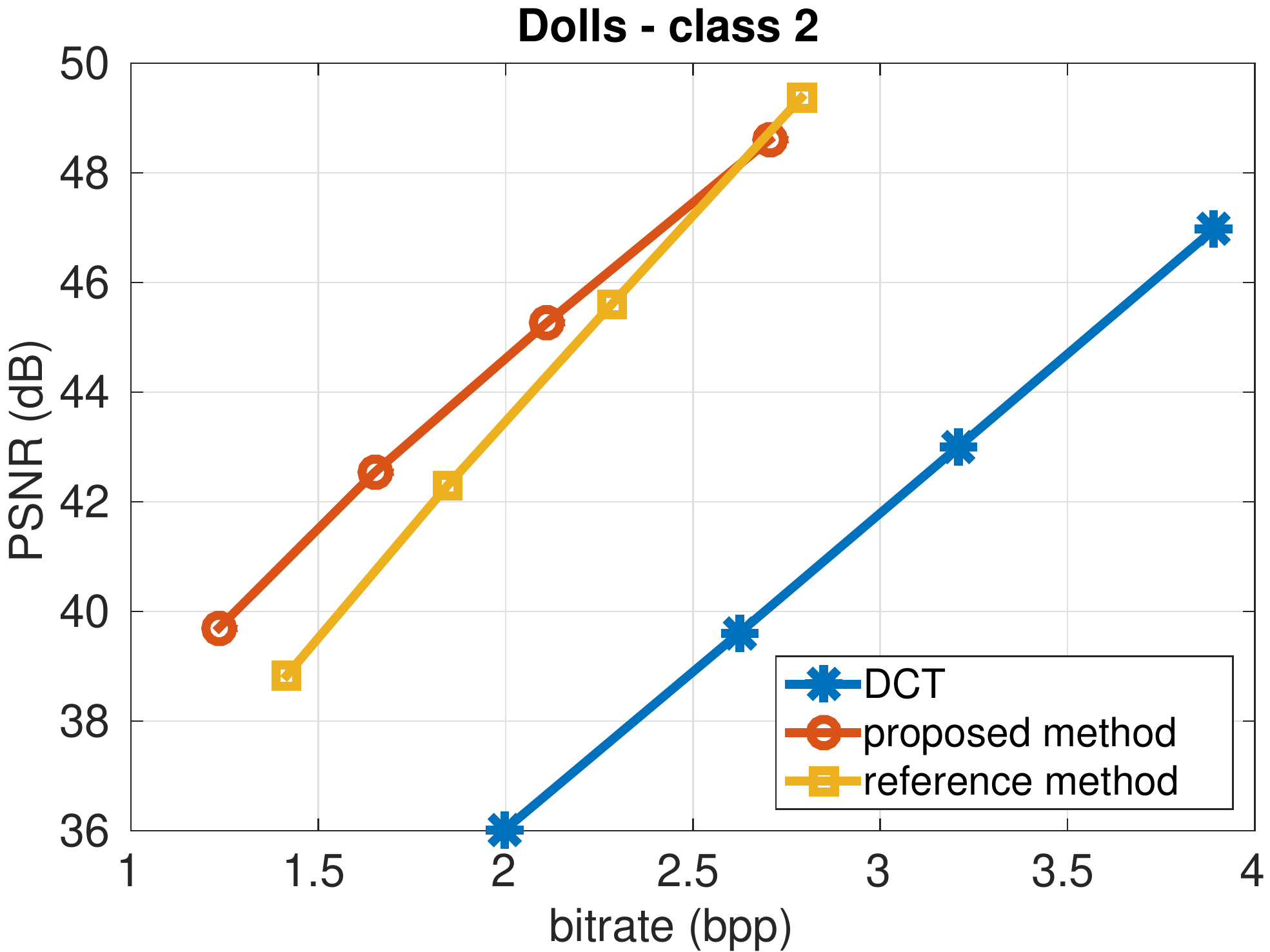}}
   \end{minipage}
   \vfill
\begin{minipage}[b]{\linewidth}
  \centering
  \centerline{\includegraphics[width=7cm]{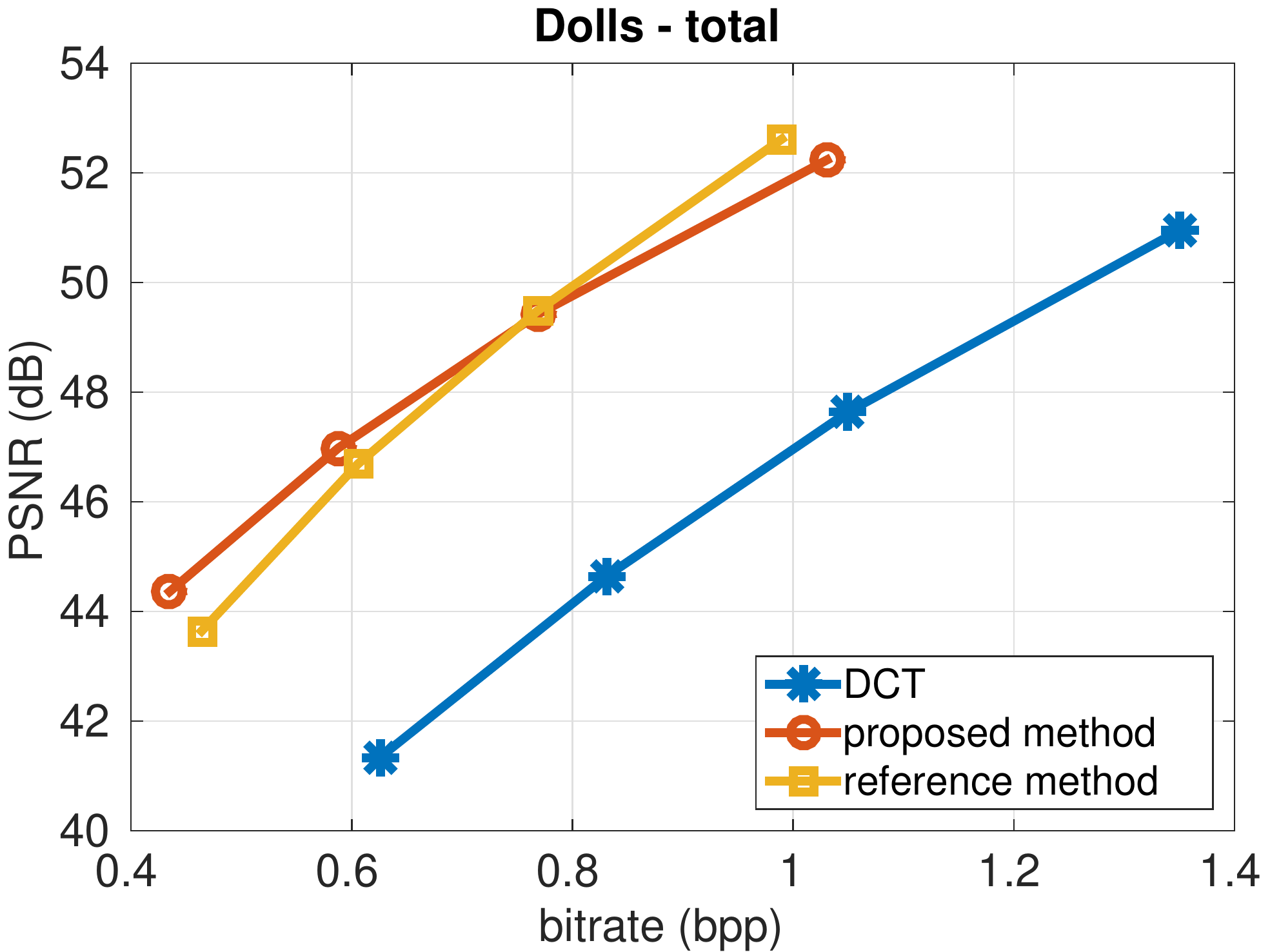}}
   \end{minipage}
 \caption{Rate-distortion curves for the image Dolls. }
\label{fig:rd_depth}
\end{figure}

\begin{figure}[tb]
\minipage{\linewidth}
\centering
  \includegraphics[width=0.8\linewidth]{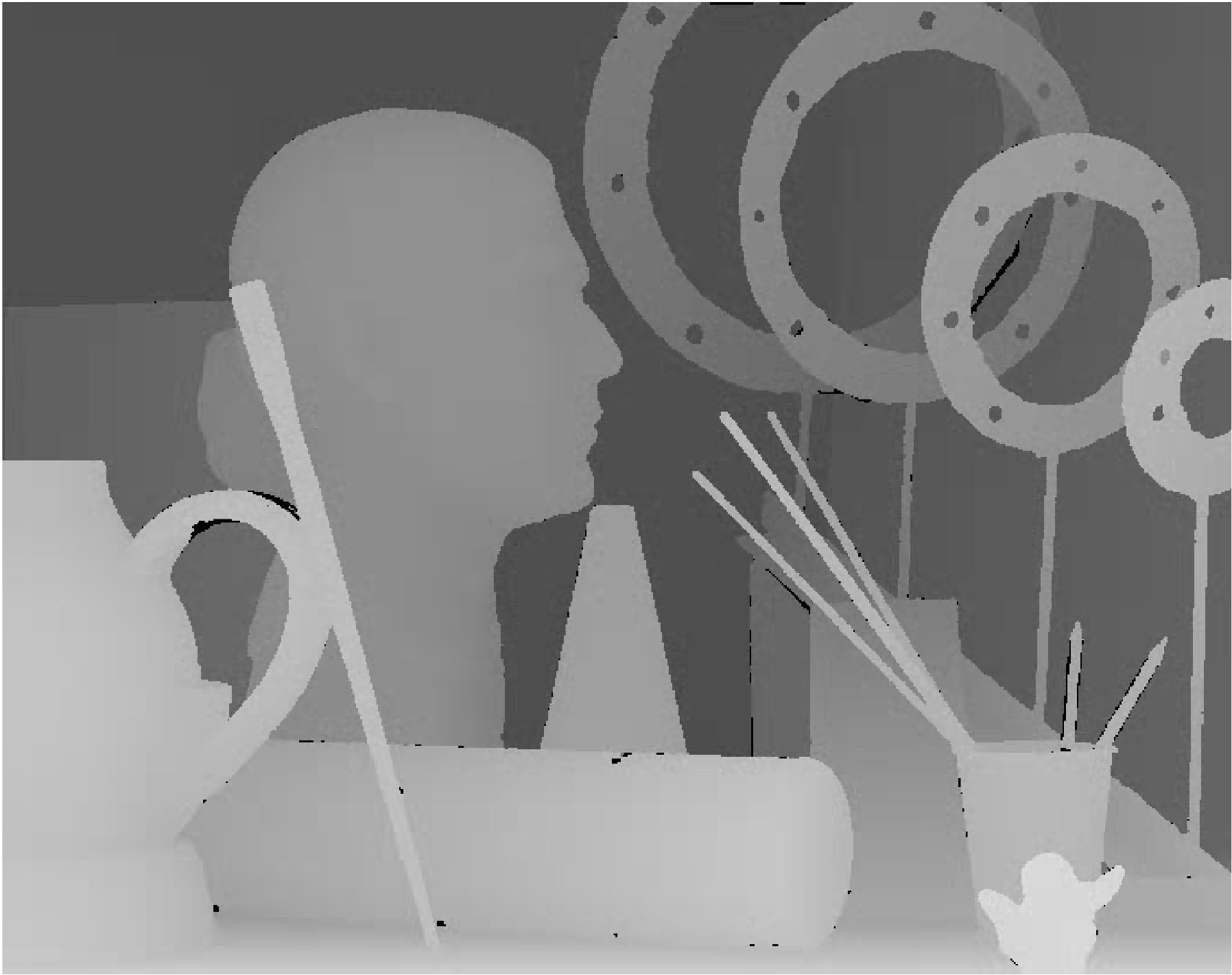}
\endminipage
\vspace{0.4cm}
\minipage{\linewidth}
\centering
  \includegraphics[width=0.8\linewidth]{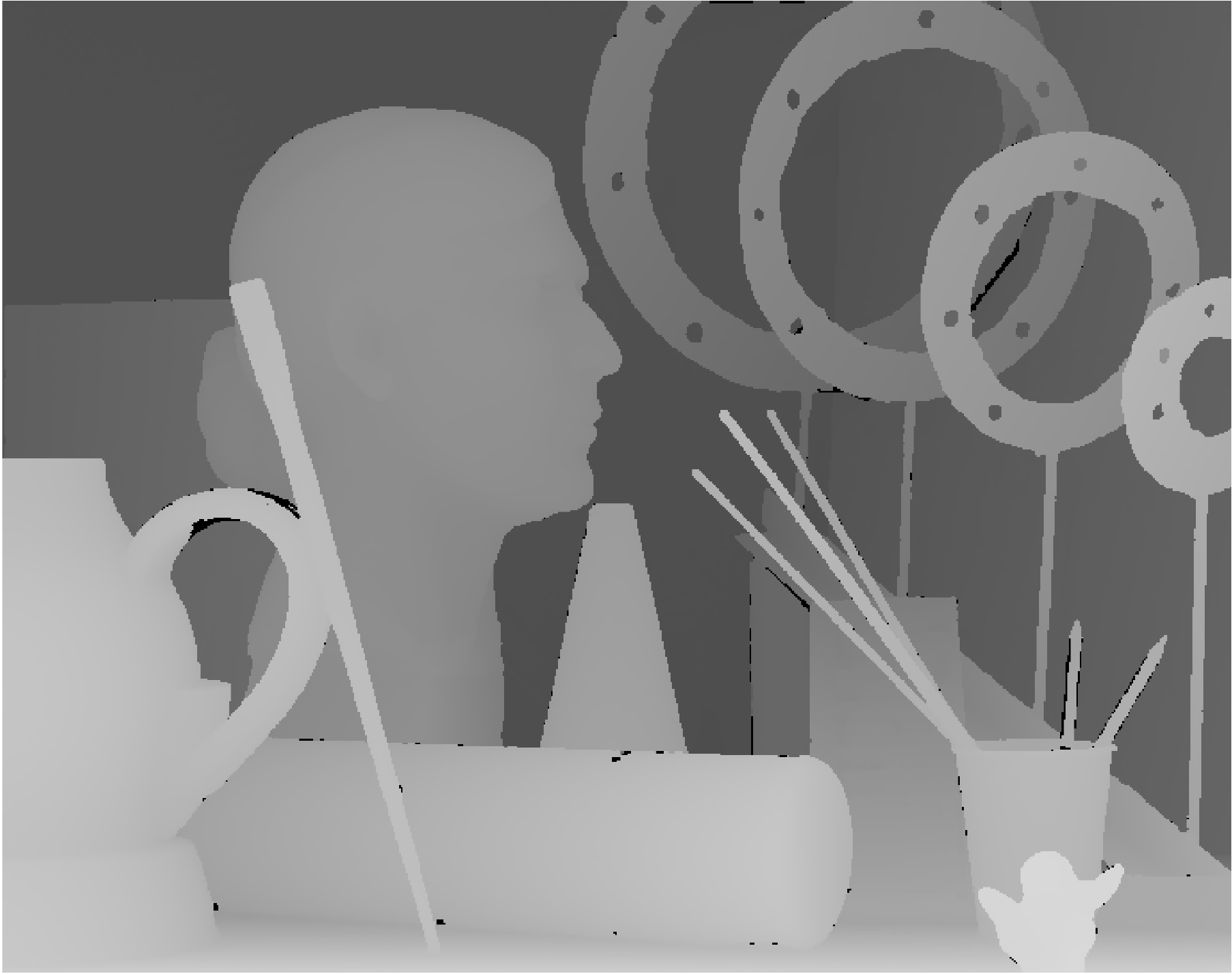}
\endminipage
 \caption{Top: Original image. Bottom: Decoded image using the proposed method (0.6 bpp).}
\label{fig:vis_depth}
\end{figure}

The first coding results on depth maps are summarized in Table \ref{tab:bjon_pws}, where we show the average gain in PSNR compared to DCT. Instead, in Table \ref{tab:bjon_pws_comp} we show the Bjontegaard average gain in PSNR between the proposed method and the reference method described previously. Moreover, in Fig. \ref{fig:rd_depth}  we show the rate-distortion curves for the image Dolls. Finally, Fig. \ref{fig:vis_depth} shows an example of a decoded image obtained using the proposed method.

The results show that the proposed technique provides a significant quality gain compared to DCT, displaying a behavior similar to other graph-based techniques. Moreover, it is important to highlight that the performance of the proposed method are close to that of the state-of-the-art method \cite{hu2015multiresolution}, although our method is not optimized for piecewise smooth images, but it is a more general method that can be applied to a variety of signal classes. In particular, for the blocks belonging to the second class, in 8 out of 10 images (namely Rocks, Cloth, Baby, Aloe, Cones, Art, Dolls and Reindeer) we are able to outperform the reference method, reaching in some cases a quality gain larger than 1 dB (see Table \ref{tab:bjon_pws_comp}). Overall, with our more generic compression framework, we outperform the reference method in more than half of the test images. In general, we observe that the proposed method outperforms the reference one in blocks that have several edges or edges that are not straight. This is probably due to the fact that, in these cases, it is more difficult to represent the graph using a lookup table. It is also worth noting that our method shows better performance at low bitrate, as it is possible to see in Fig. \ref{fig:rd_depth}.

\begin{figure}[t]
\vspace{-0.3cm}
\centering
\includegraphics[width=9cm]{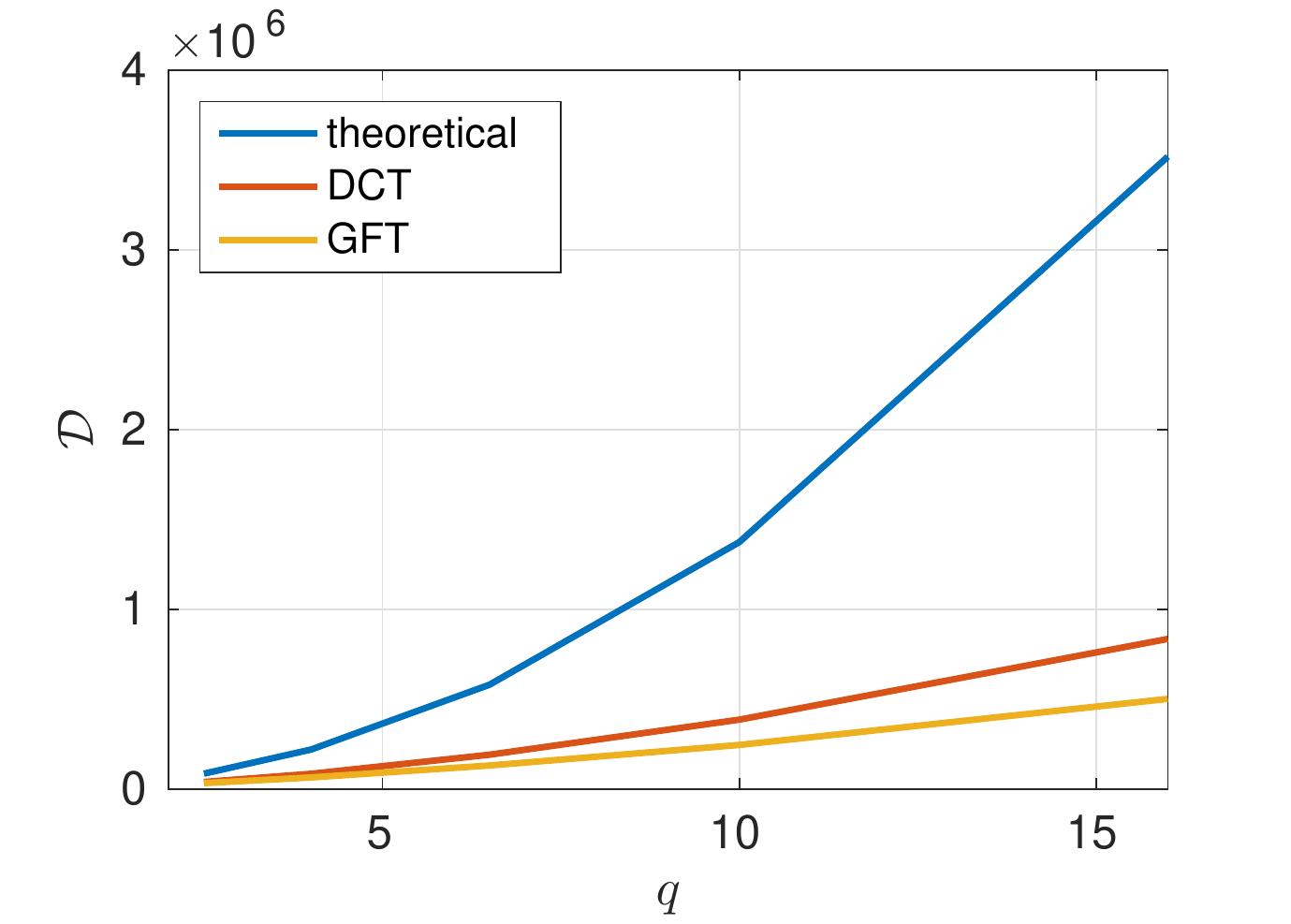}
\caption{Theoretical vs actual distortion with respect to the quantization step size $q$ for the image Teddy. The corresponding DCT and GFT bitrates lie approximately between 0.7 bpp (for $q=16$) and 1.8 bpp (for $q=2.5$) for DCT, and 0.6 bpp (for $q=16$) and 1.5 bpp (for $q=2.5$) for GFT. }
\label{fig:teo1_pws}
\end{figure}

As done in Section \ref{sec:res_val} for natural images, we also perform a validation of the theoretical assumptions used to derive the graph optimization problem \eqref{problem_final} for piecewise smooth images. We first examine the approximation of the distortion presented in Section IV.A. Fig. \ref{fig:teo1_pws} shows the theoretical and actual distortion with respect to the quantization step size $q$. We can observe that, at high bitrate (i.e., small quantization step size), the theoretical distortion converges to the actual one. Table \ref{tab:corr2} shows the correlation between the actual bitrate and the theoretical bitrate computed using \eqref{R_c} and \eqref{R_G}. The results show that \eqref{R_c} and \eqref{R_G} provide a good estimate of the actual bitrate for piecewise smooth images.

\begin{table}[t]
\centering
\caption{Correlation coefficient between theoretical and actual bitrates.}
\renewcommand{\arraystretch}{1.1}
\begin{tabular}{|c|c|c|c|c|c|}
\hline
Image        &Teddy &Cones & Art &Reindeer&Aloe\\
\hline
Correlation      &0.830&0.983&0.996&0.984&0.991\\
\hline
\end{tabular}
\label{tab:corr2}
\end{table}
\section{Conclusion}
In this paper, we have introduced a new graph construction problem targeted for compression. The solution of the proposed problem is a graph that provides an effective tradeoff between the energy compaction of the transform and the cost of the graph description. Then, we have also proposed an innovative method for coding the graph by treating the edge weights as a new signal that lies on the dual graph.   
We have tested our method on natural images and on depth maps. The experimental results show that the proposed graph transform outperforms classical DCT and, in the case of depth map coding, even compares to the state-of-the-art graph-based coding method.

We believe that the proposed technique participates to opening a new research direction in graph-based image compression.  As future work, it would be interesting to investigate other possible representation for the edge weights of the graph, such as graph dictionaries or graph wavelets. This may lead to further improvements in the coding performance of the proposed method. 

\section*{Acknowledgment}
This work was partially supported by Sisvel Technology.

\bibliographystyle{IEEEtran}

\end{document}